\documentclass[reprint,amsmath,amssymb,aps,prb,groupedaddress,nofootinbib,twocolumn,superscriptaddress]{revtex4-2} 
\usepackage{graphicx}
\usepackage{amsthm,amssymb,amsmath,braket,mathdots}
\usepackage{bm}
\usepackage{physics}
\usepackage[pagebackref=false,pdfnewwindow=true]{hyperref} 
\usepackage{epstopdf,psfrag}
\usepackage{relsize,amsbsy}
\usepackage[export]{adjustbox}
\usepackage{stmaryrd}

\usepackage{graphicx,xcolor} 
\usepackage{tabularx, booktabs} 

\usepackage{units} 
\usepackage{dsfont} 

\renewcommand{\vec}[1]{\bm{#1}} 

\renewcommand{\eqref}[1]{Eq.~(\ref{#1})}

\usepackage{tikz}
\usetikzlibrary{arrows.meta}

\definecolor{bananayellow}{rgb}{1.0, 0.88, 0.21}
\definecolor{straw}{rgb}{0.32, 0.28, 0.1}

\usepackage[caption=false]{subfig}
\usepackage{changes}
\definechangesauthor[name=Ignacio, color=blue]{ISL} 
\definechangesauthor[name=Johannes, color=red]{JK}
\definechangesauthor[name=Alex, color=green]{AM} 
\definechangesauthor[name=Leo, color=magenta]{LM}

\begin{document}

\title{Thermal Hall conductivity from semiclassical spin dynamics simulations: implementation and applications to chiral ferromagnets and Kitaev magnets}
\author{Ignacio Salgado-Linares}
\affiliation{Technical University of Munich, TUM School of Natural Sciences, Physics Department, TQM, 85748 Garching, Germany}
\affiliation{Munich Center for Quantum Science and Technology (MCQST), Schellingstr. 4, 80799 M{\"u}nchen, Germany}
\author{Alexander Mook}
\affiliation{Institute of Solid State Theory, University of Münster, D-48149 Münster, Germany}
\author{Léo Mangeolle}
\affiliation{Technical University of Munich, TUM School of Natural Sciences, Physics Department, TQM, 85748 Garching, Germany}
\affiliation{Munich Center for Quantum Science and Technology (MCQST), Schellingstr. 4, 80799 M{\"u}nchen, Germany}
\author{Johannes Knolle}
\affiliation{Technical University of Munich, TUM School of Natural Sciences, Physics Department, TQM, 85748 Garching, Germany}
\affiliation{Munich Center for Quantum Science and Technology (MCQST), Schellingstr. 4, 80799 M{\"u}nchen, Germany}
\affiliation{Blackett Laboratory, Imperial College London, London SW7 2AZ, United Kingdom}
\date{\today}

\begin{abstract}
We investigate thermal Hall transport in magnetic systems, using semiclassical spin dynamics simulations. Building on a linear response framework, we discuss the intricacies of computing the thermal Hall conductivity from real-time energy current correlations and the energy magnetization. We then apply this methodology to two models: a square-lattice chiral magnet with in-plane Dzyaloshinskii–Moriya interaction, and the antiferromagnetic Kitaev model in a field. Our results demonstrate the efficiency of semiclassical spin dynamics to study thermal Hall transport capturing quantitative effects beyond the simple intrinsic non-interacting approximation. They can serve as a benchmark for comparison with experiments in regimes where non-linearities from magnon-magnon interactions and strong thermal fluctuations play a crucial role.
\end{abstract}
	
\maketitle

\section{Introduction}
\label{sec:intro}

The thermal Hall conductivity is a fundamental, yet still only partly understood, property of solid state materials, that has attracted considerable attention in the past few years. In particular, it has been proposed as a signature of exotic excitations in quantum materials, as the sought-for half-quantization of $\kappa_{xy}/T$ would be a sharp signature of fractionalization, and has been predicted in fractional quantum Hall states \cite{kane1997quantized}, in topological superconductors~\cite{sumiyoshi2013quantum} and in some fermionic spin liquids, notably Kitaev's honeycomb model, where (in the gapped phase at $T=0$) it is carried by chiral Majorana fermions~\cite{kitaev2006anyons}. The thermal Hall effect (THE) is particularly promising for probing intrinsic properties of charge neutral quasiparticles in insulating quantum magnets~\cite{zhang2024thermal}. However, over the last years it has been realized that a simple interpretation of experimentally observed THEs in terms of intrinsic Berry curvature contribution of quasiparticles, e.g. magnons~\cite{matsumoto_rotational_2011}, is challenging. One reason is that phonon thermal transport always contributes in addition to any magnetic contribution, which can spoil quantized THE signatures~\cite{ye2018quantization,vinkler2018approximately}. Besides, finite temperature fluctuations \cite{joy2022}, extrinsic contributions~\cite{side-jump} and intrinsic non-linear effects~\cite{dimos2025thermal} can dominate. Thus, on the experimental side, identifying the origin of a THE, both regarding its carriers (phonons, magnons, fractionalized excitations) and its mechanism (intrinsic, extrinsic), remains an open challenge.

The THE is notoriously hard to tackle analytically, especially insofar as many species of energy carriers may be involved. While many studies, in magnetic materials, focus on the intrinsic contribution from non-interacting magnons in clean systems (see [see Ref. \cite{matsumoto_rotational_2011} or \eqref{eq:kappaxy} below), which only requires knowledge of the magnon dispersion and Berry curvature in the harmonic approximation, the regime of validity of this picture remains unclear. Beyond this intrinsic mechanism, the physics of THEs is presumably much richer, including in particular many-body skew-scattering between magnons \cite{dimos2025thermal} or involving different types of excitations \cite{mangeolle2022phonon}, impurity-induced side-jump effects \cite{side-jump}, etc. Although the intrinsic regime is often assumed, there is little evidence that it can reproduce experimental observations, either quantitatively or even often qualitatively. A more complete theoretical description, potentially relying on unbiased numerical methods, is \emph{a priori} required. Of course, as an out-of-equilibrium, dynamical, and finite-temperature property, the THE is itself highly challenging to access numerically, since reliable estimates demand large system sizes whose time evolution and thermalization become increasingly costly.

Here, we present the numerical study of the THE in magnetic systems using \textit{semiclassical spin evolution}. This evolution relies on a coherent-state representation of spins, whose effective dynamics is essentially classical and described by a Landau-Lifshitz equation. Such method allows to consider systems of comparatively large size, and has proven successful in evaluating dynamical quantities.
One key advantage over other methods is that they are unbiased toward one type of energy carrier or of mechanism, and may capture various contributions from more or less well-defined excitations on an equal footing, including all nonlinear effects.
The use of such methods to evaluate thermal conductivity in materials has a long history. 
In magnetic systems, efforts have been made to incorporate spin dynamics to compute longitudinal (in particular, thermal) transport \cite{PhysRevB.75.214305,PhysRevB.92.134305,PhysRevB.100.144416,PhysRevB.106.224407,PhysRevB.105.104405}. The calculation of thermal Hall conductivity, which requires the breaking of time reversal symmetry, poses more challenges.  
Pioneering works on magnon thermal Hall transport showed that the thermal Hall conductivity can be extracted from semiclassical dynamics \cite{mook_spin_2016,PhysRevB.95.020401} but neglected the contribution of energy magnetization. Despite a lot of interest in THE of quantum magnets, only very few calculations based on semi-classical spin dynamics have been reported, i.e. for a chiral ferromagnetic model on the square \cite{carnahan_thermal_2021}, and the triangular lattice \cite{Kim2024}.

Here, we generalize this procedure and describe in detail the technical challenges and implementation steps, that can be generalized to (in principle) any spin model on any lattice. We use the results of Ref.\cite{carnahan_thermal_2021} as a benchmark and discuss the convergence properties in terms of system size and equilibration time scales. We then explore unchartered territory investigating the antiferromagnetic Kitaev model just above its saturation field. In the polarized phase near the predicted (classical within our description) spin liquid phase, it is not clear a priori whether coherent intrinsic Hall transport from well-defined magnon excitations, or a thermal Hall signature reminiscent of the predicted half-quantized value from fractionalized excitations in the corresponding quantum spin liquid phase, should be expected. This motivates our investigation using unbiased semiclassical numerics. We find general agreement with magnonic transport, yet a failure of the intrinsic LSWT formula for the THE -- see our Fig.\ref{Fig06}. Indeed, we show that thermal fluctuations and nonlinearities have a marked effect on the thermal Hall conductivity, not only because they renormalize intrinsic energy-carrying quasiparticles \cite{koyama2024thermal}, but also because they can generate other mechanisms for THE.

\begin{figure}[t]
    \centering
    \includegraphics[width=\columnwidth]{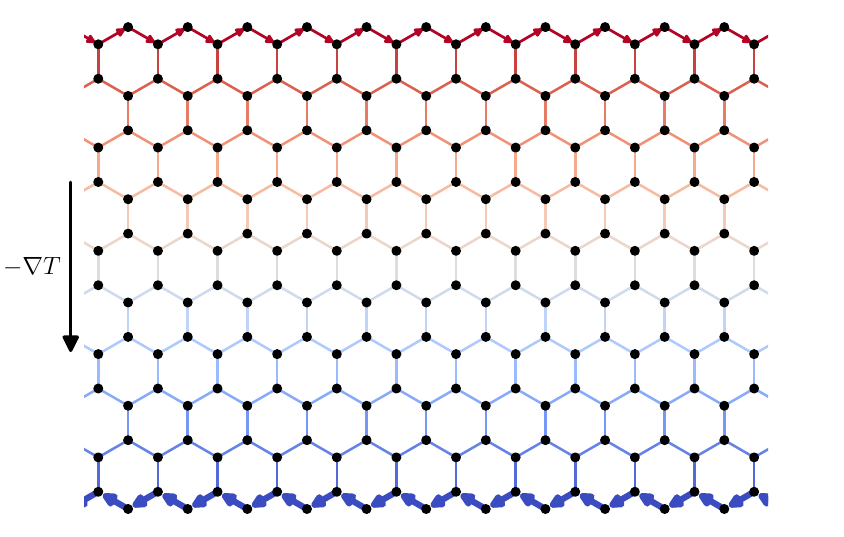}
    \caption{Schematic representation of local equilibrium (i.e.~magnetization) thermal Hall currents in a honeycomb lattice with a temperature gradient $\nabla T$. The lattice bonds are colored according to the local temperature. On each bond, an arrow represents the local energy current, whose width encodes the current's magnitude. In the systems we consider, we find that at local equilibrium, only chiral edge energy currents are sizable, however bulk currents remain crucial to determine the response out of equilibrium.}
    \label{fig:nicefigureone}
\end{figure}

Our paper thus serves two main purposes. First, we aim at providing a detailed, benchmarked, pedagogical exposition of how to numerically extract the thermal Hall conductivity from classical spin dynamics simulations. Second, by applying this method to the antiferromagnetic Kitaev model in the polarized phase proximate to the spin liquid, we want to investigate the suitability of semiclassical methods in fluctuating regimes. We thus establish that they provide physically meaningful results and show that the simple interpretation of the THE in terms of intrinsic Berry phase effects of harmonic spin waves is insufficient.

    The rest of the paper is organized as follows. In section \ref{sec:theory}, we present the theoretical background for our numerical evaluation of thermal Hall conductivity for spin systems, and for comparison with other approaches. In section \ref{sec:numerical_methods}, we expose in detail our numerical setup for the semiclassical spin evolution and extraction of current-current correlations and equilibrium statistical averages. In section \ref{sec:results} we apply this procedure to two physical models: after benchmarking our method (Sec. \ref{sec:chiral_fm}), we investigate Kitaev's honeycomb model in a field (Sec. \ref{sec:kitaev_afm}), then we discuss the relation between our approach and other methods (Sec. \ref{sec/discussion}).

\section{Theory}
\label{sec:theory}

\subsection{Defining local energy currents}
\label{sec:defloccur}

We begin our discussion by presenting the theoretical framework used to compute the thermal Hall conductivity within a semi-classical formalism, based on linear response theory applied to energy currents described in terms of classical spin variables \cite{PhysRevB.91.125413}. The Hamiltonian, in any local theory, can be constructed as the sum of local energy densities, $\mathcal{H}=\sum_i h_i$. Then, a local energy current can be defined, either locally from the continuity equation (as an operator whose lattice divergence is $-\partial_t h_i$) or from a more global perspective via the energy polarization \cite{nomura2012cross}. The latter can be defined as
\begin{equation}
    \boldsymbol{P}_E \equiv \frac{1}{V} \sum_i \boldsymbol{r}_i h_i,
\end{equation}
where $V$ is the volume of the system and $\boldsymbol{r}_i$ the coordinate vector of site $i$ on the lattice. The time derivative of this quantity yields the (global) energy current density, 
\begin{equation}
    \boldsymbol{I} =  \frac{1}{V} \sum_i \boldsymbol{r}_i \dot{h}_i.
\end{equation}
Here we note that the sum $\sum_i$ can either be over the whole system, or over any mesoscopic region, since $\boldsymbol{P}_E$ and $\boldsymbol{I}$ can equivalently be defined not globally but on a coarse-grained scale, leaving the physics unchanged.

Now we assume that each operator $h_i$ is given in terms of products of local spin variables $\boldsymbol{S}_k$ pertaining to arbitrary lattice sites. To compute the time derivative $\dot h_i$, we assume the dynamics of spins is \emph{semi-classical}, so that the local energy density operator $h_i$ evolves under Hamilton's equation $\dot{h}_i=\{h_i, \mathcal{H}\}$ where $\{\cdot,\cdot\}$ is the Poisson bracket. This allows us to rewrite the current density as $\boldsymbol{I}=\frac{1}{V} \sum_i \sum_j \boldsymbol{r}_i \{h_i, h_j\}$. We use the identity $\{f,g\}=-\{g,f\}$ to symmetrize the sum, yielding
\begin{equation}\label{eq:current_0} 
    \boldsymbol{I}=\frac{1}{V} \sum_i \sum_{j<i} (\boldsymbol{r}_i-\boldsymbol{r}_j)\{h_i, h_j\}
    = \frac{1}{V} \sum_i \sum_{j<i} \boldsymbol j_{j \rightarrow i},
\end{equation}
where $\boldsymbol j_{j \rightarrow i}=(\boldsymbol{r}_i-\boldsymbol{r}_j)\{h_i, h_j\}$ is the local bond current,
which identifies as the heat current from site $j$ to site $i$. 
Here $j<i$ refers to some arbitrary ordering of sites which avoids double-counting. In general the summation also involves not only nearest-neighbors but also further neighbors (namely distant by at most twice the range of interactions).

In this section we have defined local energy currents generally for any local hamiltonian $h_i$ regardless of which degrees of freedom it involves. Just $\mathcal H = \sum_i h_i$ and \eqref{eq:current_0} are sufficient to derive general formulas for the thermal Hall conductivity, Sec. \ref{sec:lrt_hall}. We will address the particular case of spin models, where $h_i$ consists of combinations of local spin operators, in Sec. \ref{sec:energy_currents}.

\subsection{Linear response theory for thermal Hall conductivity}
\label{sec:lrt_hall}

Having defined local energy current operators, we now seek to evaluate their non-equilibrium average value induced by a thermal gradient. For this, we employ linear response theory, originally developed by Luttinger \cite{luttinger_theory_1964} in the time-reversal (TR) invariant case, then expanded by several authors over decades \cite{Smrcka_1977, cooper1997, qinniushi2011, thermoEM2020} to describe Hall currents in the TR-broken case. Because there are physical subtleties in this construction that need to be carefully considered in the numerics, we reproduce below some key features of the derivation (see e.g.~Ref.\ \cite{carnahan2022spin} for a pedagogical exposition).

We begin by introducing a pseudo-gravitational potential $\psi_i$, which couples to the local energy density as $h_i \rightarrow h_i'$,
\begin{equation}
    h_i^\prime = h_i \left ( 1 + \psi_i \right ) .
\end{equation}

The global and local energy currents are then modified as follows:
\begin{equation}
\label{eq:5}
    \begin{split}
        \boldsymbol{I}' &=\frac{1}{V} \sum_i \sum_{j<i} (\boldsymbol{r}_i-\boldsymbol{r}_j) \{h_i^\prime,h_j^\prime\} \\
        &= \frac{1}{V} \sum_i \sum_{j<i} (\boldsymbol{r}_i-\boldsymbol{r}_j) (1+\psi_i +\psi_j) \{h_i,h_j\} \\
        &= \frac{1}{V} \sum_i \sum_{j<i} (1+\psi_i + \psi_j) \,\boldsymbol{j}_{j \rightarrow i} ,
    \end{split}
\end{equation}
to linear order. 

As mentioned above, for any local Hamiltonian (e.g.\ a spin model with short-ranged magnetic exchange) the summation involves sites $(i,j)$ that are relatively close neighbors, so one can approximate $\psi_i \approx \psi_j$ in \eqref{eq:5}. Further assuming a locally uniform gradient $\psi_i=\boldsymbol r_i \cdot \grad \psi$ and Luttinger's prescription $\grad \psi={\grad T}/{T}$, we obtain 
\begin{align}
    \boldsymbol{I}' & =\frac{1}{V} \sum_i \sum_{j<i} \boldsymbol{j}_{j \rightarrow i} 
    \,+\,\frac{2}{T}\frac 1 V  \sum_i (\boldsymbol{r}_i \cdot {\boldsymbol \nabla} T)
    \sum_{j<i} \boldsymbol{j}_{j \rightarrow i} \nonumber \\
    & = \boldsymbol{I} + \boldsymbol{\delta \! I}.
\end{align}
The energy current density operator splits into two components: an unperturbed current operator $\boldsymbol{I}$, identical to \eqref{eq:current_0}, and a linear perturbation $\boldsymbol{\delta \! I}$. 

Our interest lies in the transverse component of the conductivity, characterized by $\left\langle I'_x \right\rangle =-\kappa_{xy} \nabla_y T$,
where $\kappa_{xy}$ is the thermal Hall conductivity. The Hall conductivity decomposes into two components:
 \begin{equation}\label{eq:kappa_tot}
    \kappa_{xy}=-\frac{1}{\nabla_y T} \big (\langle I^x\rangle ' + \langle \delta \! I^x\rangle \big )= \kappa^{xy}_{\rm Kubo} + \kappa^{xy}_{\rm EM},
 \end{equation}
where the $\langle \cdot \rangle '$ statistical average needs to be performed using the \emph{non-equilibrium} (perturbed) density matrix, while $\langle \cdot \rangle $ involves only the equilibrium density matrix.

The first term is the usual contribution from Kubo's formula,
 \begin{equation}\label{eq:kappa_0}
    \begin{split}
    \kappa^{xy}_{\rm Kubo} = \frac{V}{k_{\rm B} T^2} \int_{0}^{\infty} dt \left\langle I^x(t) I^y(0) \right\rangle,
    \end{split}
 \end{equation}
with $k_{\mathrm{B}}$ the Boltzmann's constant, obtained by integrating the $I^x$ operator against the linear-in-$\nabla_y T$ contribution to the density matrix.
The second term is
 \begin{equation}\label{eq:kappa_1}
   \kappa^{xy}_{\rm EM} = -\frac{2}{T}\frac 1 V  \sum_i \sum_{j<i} \left\langle r^{y}_{i} \,j_{j \rightarrow i}^{x} \right\rangle .
 \end{equation}
In both \eqref{eq:kappa_0} and \eqref{eq:kappa_1},
the average now involves the equilibrium density matrix.

The first contribution \eqref{eq:kappa_0} represents the bulk thermal transport arising from the out-of-equilibrium response of the system to a thermal gradient, resulting in a net transverse energy flow provided TR is broken. By contrast, \eqref{eq:kappa_1} arises from magnetization currents, that are circulating even in thermal equilibrium but yielding a finite contribution to the transport current only once temperature imbalance is introduced. 

Crucially, only the sum of the Kubo term and the energy magnetization term 
is a physically meaningful quantity, in the sense that it is rid of a spurious $1/T$ divergence for gapless free bosons (see Sec. V in Ref. \cite{qin_em})
and that it satisfies locally the third law of thermodynamics \cite{starpaper}.

Alternatively,the magnetization contribution can also be formulated in momentum space in the spirit of linear response \cite{qinniushi2011, thermoEM2020, Kim2024}. 
It then reads
\begin{equation}\label{eq:kappa_1m}
\kappa^{xy}_{\text{EM}} = \frac{2 M_{\rm E}^{z}}{V T},
\end{equation}
with $M_{\rm E}^{z}$ the $z$-component of the energy magnetization $\bm M_{\rm E}$. 
The latter may be computed, as linear response determines its derivative with respect to a thermodynamical conjugate variable $(\mu,T,P,\dots)$ -- then $\bm M_{\rm E}$ can be obtained upon integration. Assuming it vanishes in the high-temperature limit, 
its $z$-component is given by
\begin{equation}
    M_{\rm E}^{z}(T) = T^2 \int_T^\infty \frac{\mu_{\rm E}^{z}(T’)}{T’^3}  dT’,
\end{equation}
where the integrand involves the energy magnetization density $\mu_{\rm E}^{z}(T')$ defined as
\begin{equation}
\label{eq:muE}
\boldsymbol{\mu}_{\rm E} = \frac{1}{2 i T} \nabla_{\mathbf{k}} \times \left\langle h_{-\mathbf{k}} ; \boldsymbol{j}_{\mathbf{k}} \right\rangle \Big|_{\mathbf{k} \to 0},
\end{equation}
with $h_{-\mathbf{k}} = \sum_i e^{-\mathrm{i} \mathbf{k} \cdot \mathbf{r}_i} h_{i}$ and $\boldsymbol{j}_{\mathbf{k}} = \sum_i e^{\mathrm{i} \mathbf{k} \cdot \mathbf{r}_i} \sum_{j < i}\boldsymbol{j}_{j \rightarrow i}$ the Fourier transformed local energy density and current at each site, respectively, and 
\begin{align}
\label{eq:kubocorr}
    \left\langle h_{-\mathbf{k}} \, ; \boldsymbol j_{\mathbf{k}}  \right \rangle
    &= T \int_0^{1/T}\text d \lambda \left \langle e^{\lambda \hbar \mathcal H} h_{-\mathbf{k}} e^{-\lambda \hbar \mathcal H} \boldsymbol j_{\mathbf{k}} \right \rangle .\end{align}
In \eqref{eq:kubocorr} all objects $h_{\mathbf k},\boldsymbol j_{\mathbf k},\mathcal H$ must be understood as operators that typically do not commute, yet in the classical limit $\hbar/T \rightarrow 0$, it is a standard result (see e.g.~Ref.~\cite{pottier2009nonequilibrium}) that the quantum Kubo correlation coincides with the classical one: 
\begin{equation}
    \left\langle h_{-\mathbf{k}} \, ; \boldsymbol j_{\mathbf{k}}  \right \rangle \overset{\hbar/T \rightarrow 0}= \left\langle h_{-\mathbf{k}} \,\boldsymbol j_{\mathbf{k}}  \right \rangle .
\end{equation}
Since we assume classical spin dynamics and purely thermal fluctuations, this is the formula we use in the following.
The expression \eqref{eq:muE} captures the circulating character of the local energy magnetization currents in terms of a momentum-space torque in the long-wavelength limit. The contribution to the Hall conductivity from this circulating current is \eqref{eq:kappa_1m}.
This provides an alternative route to evaluating the energy magnetization contribution.
In the following, we will evaluate both \eqref{eq:kappa_1} and \eqref{eq:kappa_1m} to check the consistency of our numerical calculations.

We note that this decomposition of the thermal Hall conductivity into \eqref{eq:kappa_0}+\eqref{eq:kappa_1} was implemented numerically in Ref.~\cite{carnahan_thermal_2021}, albeit with a different definition of local current operators $\boldsymbol{j}_{j \rightarrow i}$ derived directly from the local continuity equation. As an early technical remark on numerical implementation, we also note that upon rescaling $\boldsymbol{j} \mapsto \zeta \boldsymbol{j}$, with $\zeta \in \mathbb R$, of the current densities, $\kappa^{xy}_{\rm Kubo} \mapsto \zeta^2 \kappa^{xy}_{\rm Kubo}$ whereas $\kappa^{xy}_{\rm EM} \mapsto \zeta \kappa^{xy}_{\rm EM}$. Therefore proper accounting of volume factors is crucial not only to correctly evaluate $\kappa_{xy}$ quantitatively, but even \emph{qualitatively} especially at low temperatures.

Finally, the longitudinal thermal conductivity
\begin{equation}
\kappa_{xx} = \frac{V}{k_{\rm B} T^2} \int_{0}^{\infty} dt \left\langle I^x(t) I^x(0) \right\rangle
\end{equation}
is unaffected by magnetization currents and given purely in terms of Kubo's usual current-current correlation.

\subsection{Intrinsic thermal Hall conductivity within linear spin-wave theory (LSWT)}
\label{sec:intrinsic-lsw}

In the following sections we will compare our numerical results with analytical predictions in regimes where a general formula exists, namely that of intrinsic thermal Hall transport by well-defined non-interacting quasiparticles. In polarized phases of magnetic systems these will be bosonic magnons, i.e.\ the fundamental excitations of LSWT.
The spin wave expansion amounts to mapping spin operators $\boldsymbol S_i$ to bosonic creation and annihilation operators $(b_i^\dagger,b_i)$ then expanding the Hamiltonian in powers of these operators. To quadratic order, it always takes the Bogoliubov-de Gennes form in momentum space,
\begin{equation}
\label{eq:hbdg}
    \mathcal{H}_{\rm BdG} = \frac 1 2 \int_{\bm k} \left ( b^\dagger_{\bm k} \; b_{-\bm k} \right ) \, H_{\bm k} \, \begin{pmatrix}
        b_{\bm k} \\  b^\dagger_{- \bm k} 
    \end{pmatrix} ,
\end{equation}
to be understood as a 2-by-2 block structure.
The energies $\epsilon_{n\bm k}$ and wavefunctions $\ket
{\boldsymbol{\Psi}_{n\boldsymbol{k}}}$ of the magnon bands are then obtained as the positive eigenvalues and the corresponding eigenvectors of the dynamical matrix $D_{\bm k}= \sigma^z  H_{\bm k}$, which for $H_{\bm k}$ a $2N\times 2N$ matrix yields $N$ bands spanned by index $n \in [ \! [ 1, N ] \! ] $. 
The Berry curvature of the $n$th magnon band can be calculated using 
\begin{equation}
\label{eq:bcurv}
 \Omega_n^{z}(\bm k)=-2\,\mathrm{Im}\sum_{m \ne n} \frac{\bra
    {\boldsymbol{\Psi}_{n\boldsymbol{k}}} \frac{\partial D_{\boldsymbol{k}}}{\partial k_x} \ket
    {\boldsymbol{\Psi}_{m\boldsymbol{k}}}
    \bra
    {\boldsymbol{\Psi}_{m\boldsymbol{k}}} \frac{\partial D_{\boldsymbol{k}}}{\partial k_y} \ket
    {\boldsymbol{\Psi}_{n\boldsymbol{k}}}}{(\epsilon_{n\boldsymbol{k}}-\epsilon_{m\boldsymbol{k}})^2} .
\end{equation}

The intrinsic (clean) thermal Hall conductivity has a simple compact expression, valid for any theory of free bosons described by \eqref{eq:hbdg} or equivalent expressions. It can be obtained straightforwardly (although, quite technically), in the spirit of \cite{Smrcka_1977}
 and Sec.\ref{sec:lrt_hall} via linear response theory \cite{shindou2014}. It can also be derived from semiclassical dynamics, either using an edge-channel picture \textit{\`a  la} Landauer \cite{matsumoto_rotational_2011} or from an inhomogeneous kinetic theory \cite{starpaper}. All derivations lead to the same formula,
\begin{equation} \label{eq:kappaxy}
    \kappa^{xy}=-\frac{k_{\rm B}^2 T}{\hbar V} \sum_{n,\boldsymbol{k}} \Omega_n^{z} (\boldsymbol{k}) \, c_2 (\rho_n(\bm k)),
\end{equation}
where $ \Omega_n (\boldsymbol{k})$ is \eqref{eq:bcurv} and $\rho_n(\bm k) = 1/(e^{\epsilon_{n\bm k}/T}+\varsigma)$ is the local equilibrium distribution, and

\begin{equation} 
\label{eq:c2}
    c_2 (\rho)= \textstyle{\int_0^\rho} \text d r \left [ \ln \left (  \varsigma + 1/r \right) \right ]^2 ,
\end{equation}
where $\varsigma=+1,-1$ for bosons and fermions respectively. We note that in the classical limit of a Rayleigh-Jeans distribution $\rho_n(\bm k) = T/\epsilon_{n\bm k}$, one has $c_2(\rho) = 1/\rho$, and $\kappa^{xy}$ turns out to be temperature-independent.

Beyond free bosons and LSWT, apart from the natural extension of \eqref{eq:kappaxy} to include spectral broadening of quasiparticles \cite{koyama2024thermal}, many other contributions from extrinsic origin will enter, as we have discussed in the introduction. Overall, it is not so clear in which regime the intrinsic single-particle picture of \eqref{eq:kappaxy} should hold, although it has become a standard tool for studying magnonic Hall transport \cite{mcclarty2022topological,zhang2024thermal}. 

In this paper, we compare our numerical results to the prediction of \eqref{eq:kappaxy} as a measure of the importance of nonlinearities (i.e.\ magnon-magnon interactions) and thermal fluctuations in the non-equilibrium dynamics of the systems we study.
We now come back to the general construction of Sections \ref{sec:defloccur} and
\ref{sec:lrt_hall}, where nothing is assumed about the nature of energy currents $\boldsymbol j_{j \rightarrow i}$.

\subsection{Derivation of the energy currents}
\label{sec:energy_currents}

In Sections \ref{sec:defloccur} and
\ref{sec:lrt_hall}, we described a general derivation of the thermal Hall conductivity without assuming a specific form for the local energy density $h_i$. We only required that its time evolution should be classical and that it should involve only short-ranged interactions. We now specify that they are built out of local spin operators $\bm S_l$ at lattice site $l$. To compute explicitly the local bond current $\boldsymbol j_{j \rightarrow i} =(\boldsymbol{r}_i-\boldsymbol{r}_j)\{h_i, h_j\}$, we use the fact that for two functions $(h_i,h_j$) that are combinations of local spin operators $\bm S_l$, their Poisson bracket is known \cite{yang_generalizations_1980} to be
\begin{equation}
\label{eq/poissonbracket}
\{h_i,h_j\}=
    \sum_l \boldsymbol{S}_l \cdot \left( \frac{\partial{h_i}}{\partial{\boldsymbol{S}_l}} \times 
    \frac{\partial{h_j}}{\partial{\boldsymbol{S}_l}} \right).
\end{equation} 
The derivative of the local energy density with respect to a spin variable can be interpreted as an effective magnetic field. We thus define $\boldsymbol{B}_{l|i}= - \partial{h_i}/ \partial{\boldsymbol{S}_l}$, so that the local bond currents read
\begin{equation} \label{eq:1}
    \boldsymbol{j}_{j \rightarrow i}= (\boldsymbol{r}_i -\boldsymbol{r}_j) \sum_l \boldsymbol{S}_l \cdot \left( \boldsymbol{B}_{l|i} \times \boldsymbol{B}_{l|j} \right).
\end{equation}

We further assume that the local energy density involves at most spin bilinears,
\begin{equation}
\label{eq:h_i}
    h_i={S}_i^a J_{ii}^{ab} {S}_i^b +\frac{1}{2} \sum_{j \ne i} {S}_i^a J_{ij}^{ab} {S}_j^b - \boldsymbol{B}_{\rm ext} \cdot \boldsymbol{S}_i,
\end{equation}
including Zeeman coupling to an external magnetic field $\boldsymbol{B}_{\text{ext}}$ and where $J_{ij}^{ab}= J_{ji}^{ba}$ is the magnetic exchange matrix between sites $i$ and $j$ (with $a,b \in \{x,y,z\})$, with implicit summation over $a,b$. The first term of \eqref{eq:h_i} describes on-site anisotropy, and the second term describes pairwise interactions at arbitrary (but short) distances from site $i$, with a $1/2$ prefactor that avoids double-counting. 
Differentiating with respect to $\boldsymbol{S}_l$ yields
\begin{align}
\label{eq:Blia}
{B}_{l|i}^a 
&= \delta_{il}  \left ( B^a_{\rm ext} 
- 2  J_{ii}^{ab} {S}_i^b \right ) 
- \frac{1}{2} \delta_{il} \sum_{j \ne i} J_{ij}^{ab} {S}_j^b 
- \frac{1}{2} {S}_i^b J_{il}^{ba}  ,
\end{align}
where the last term interprets as the Weiss-Onsager field felt by the spin at site $l$ caused by the spin at site $i$, and summation over $b$ is implicit. \eqref{eq:Blia} can be straightforwardly implemented numerically, for arbitary magnetic exchange $J_{ij}^{ab}$.

\section{Numerical details}
\label{sec:numerical_methods}

\subsection{Classical stochastic spin dynamics}
\label{sec:spin_dyn}

We compute the spin dynamics of the system by numerically integrating the stochastic Landau-Lifshitz-Gilbert (LLG) equation
\begin{equation}\label{eq:llg}
    \frac{d\boldsymbol{S}_i}{dt}=-\frac{\gamma}{1+\alpha_{\rm G}^2} [\boldsymbol{S}_i \times \boldsymbol{B}_i^{\mathrm{eff}} + \alpha_{\rm G} \boldsymbol{S}_i \times (\boldsymbol{S}_i \times \boldsymbol{B}_i^{\mathrm{eff}})]
\end{equation}
where $\gamma$ is the gyromagnetic ratio and $\alpha_{\rm G}$ is the Gilbert damping coefficient. In \eqref{eq:llg}, the first term describes the precessional motion of local spins $\boldsymbol{S}_i$, constrained to be unit-length vectors, around the effective magnetic field $\boldsymbol{B}_i^{\mathrm{eff}}$ at site $i$. The second term is a phenomenological damping term \cite{gilbert2004phenomenological}, accounting for some memory effects induced by interactions of a local spin with the surrounding system's thermal fluctuations beyond classical approximation, which enforces the relaxation of a spin $\boldsymbol{S}_i$ to align with the local magnetic field $\boldsymbol{B}_i^{\mathrm{eff}}$.

As a classical equation of motion, \eqref{eq:llg} (without damping) neglects both quantum and thermal fluctuations. Here, thermal fluctuations are crucial to equilibrate the system in the canonical ensemble and eventually evaluate statistical averages $\langle \cdot \rangle$ at a given temperature. We account for these fluctuations by introducing Langevin dynamics in the effective magnetic field $\boldsymbol{B}_i^{\mathrm{eff}}$. We decompose $\boldsymbol{B}_i^{\mathrm{eff}}=\boldsymbol{B}_i^{\mathrm{cl}} + \boldsymbol{B}_i^{\mathrm{th}}$, where
\begin{equation}
    \boldsymbol{B}_i^{\mathrm{cl}}=-\frac{1}{\mu_{\rm B}} 
    \frac{\partial \mathcal{H}}{\partial \boldsymbol{S}_i}
\end{equation}
(with $\mu_{\rm B}$ the Bohr magneton) is the sum of the external magnetic field $\boldsymbol B_{\rm ext}$ and the Weiss-Onsager reaction field, and $\boldsymbol{B}_i^{\mathrm{th}}$ is a stochastic field. We assume the latter is a Gaussian white noise, whose correlations (between components $a,b \in \{x,y,z\}$ and sites $i,j$) are
\begin{equation}
\label{eq:BB}
  \left \langle ({B}_i^{\mathrm{th}})^a(t) \, ({B}_j^{\mathrm{th}})^b(t') \right \rangle = \frac{2 \alpha_{\rm G} k_{\mathrm{B}} T}{{\gamma \mu_{\rm B} }} \; \delta(t-t')\,\delta_{ij}\,\delta_{ab}.
\end{equation}
Because thermal fluctuations are ultimately responsible for the relaxational dynamics towards local equilibrium, they satisfy the fluctuation-dissipation theorem \cite{eriksson2017atomistic}, which relates them to the Gilbert damping coefficient $\alpha_{\rm G}$. Thus, the system's temperature $T$ is defined by the ratio of the Langevin field's variance in \eqref{eq:BB} to the damping coefficient in \eqref{eq:llg}. We note that alternatively, it is also possible to use classical Monte Carlo sampling to equilibrate the system.

\subsection{Stochastic vs Hamiltonian time evolution}
\label{sec:cano-microcano}

The stochastic time evolution described in Sec.\ref{sec:spin_dyn} allows physically to thermalize the spin system at a given temperature $T$, which permits to compute thermal averages in the canonical ensemble. It cannot, strictly speaking, be used to study the time correlators of dynamical quantities (especially energy currents derived from a continuity equation) which should be governed by \emph{Hamiltonian} evolution, in the microcanonical ensemble. Thus our numerical simulations are decomposed into two phases:
\begin{itemize}
    \item[$-$] An equilibration phase, at times $t \in [-t_{\rm eq},0]$, where the spin system evolves from an arbitrary initial configuration following the Langevin stochastic dynamics of Sec.\ref{sec:spin_dyn} with the full $\boldsymbol{B}_i^{\mathrm{eff}}=\boldsymbol{B}_i^{\mathrm{cl}} + \boldsymbol{B}_i^{\mathrm{th}}$. At the end of this phase, thermal averages of static quantities like \eqref{eq:kappa_1} can be sampled as the system has reached canonical equilibrium.
    \item[$-$] A hamiltonian evolution phase, at times $t \in [0,t_{\rm max}]$, where the spin system evolves under hamiltonian dynamics, i.e.\ \eqref{eq:llg} where $\alpha_{\rm G}=0$ (and accordingly $\boldsymbol{B}_i^{\mathrm{eff}}=\boldsymbol{B}_i^{\mathrm{cl}} + 0$). Snapshots can then be taken at any time, to evaluate thermal averages of dynamical quantities like \eqref{eq:kappa_0}.
\end{itemize}

In both phases, we compute the semiclassical spin evolution by using the \textsc{Vampire} spin dynamics package \cite{Evans2014}. The integration of the LLG equation is performed using the mid-point method \cite{ellis_midpoint_2012}, that is a second-order Runge-Kutta method. This is particularly suitable for integrating spin trajectories in a microcanonical setting, as it is a symplectic,  norm-conserving scheme that efficiently enforces energy conservation over the computational time scales \cite{daquino_symplectic, Mentink_2010}.

In fact, to evaluate the energy magnetization contribution (e.g.\ using the real space formulation of \eqref{eq:kappa_1}), instead of equilibrating the system a large number of times (each with $n_{\mathrm{eq}}$ equilibration steps), it is computationally much more efficient to only equilibrate the system once. We then keep evolving it under Langevin dynamics at later times $t \in [0,t_{\rm th}]$, taking snapshots at intervals $t_{\rm snap}$ to obtain a large statistical average with $t_{\rm th}/t_{\rm snap}$ samples. The efficiency of this procedure relies on $t_{\rm snap} \ll t_{\rm eq}$, which is guaranteed by the following (physically reasonable and numerically easy to check) fact: it generally takes much longer for a system to reach ergodicity from an arbitrary, possibly far from equilibrium, initial state, than it takes to erase memory of past states in the regime with (approximate) ergodicity {\cite{thesishellsvik, frenkel2002}.

\subsection{Extraction of thermal current-current correlations} 
\label{sec:fitting}

As recalled in section \ref{sec:lrt_hall}, the thermal Hall conductivity can be decomposed into two contributions, $\kappa_{xy}=\kappa^{xy}_{\rm Kubo} + \kappa^{xy}_{\rm EM}$. 
The magnetization contribution $\kappa^{xy}_{\rm EM}$ can be evaluated straightforwardly from \emph{static} averages, and does not pose much computational difficulty (see Sec.\ref{sec:cano-microcano}). Here we focus on the Kubo contribution $\kappa^{xy}_{\rm Kubo}$, which is hardest to evaluate properly.

The latter is extracted, using \eqref{eq:kappa_0}, from the \emph{dynamical} current-current correlation functions 
\begin{equation}
\label{eq:Cab}
    C^{\alpha\beta}(t) = \theta(t)\left\langle I^{\alpha}(t)\, I^{\beta}(0) \right\rangle ,
\end{equation}
with $\alpha,\beta \in \{x,y\}$, $\theta$ the Heaviside function, and $\langle \cdot \rangle$ is the statistical (thermal) average. This requires computing the energy currents at all times of the spin dynamics evolution, then averaging over stochastic samples.

In order to efficiently eliminate statistical noise, it is useful to fit the correlation function \eqref{eq:Cab} to the following forms,
\begin{subequations}
\label{eqs:28}
\begin{align}
    C^{xy}_{\rm fit}(t) &= \theta(t)\sum_{m=1}^{M} A_m^\perp e^{-B^\perp_m t} \sin(C^\perp_m t), \label{eq:correlation_fit_xy} \\
    C^{xx}_{\rm fit}(t) &= \theta(t)\sum_{m=1}^{M} A_m^\parallel e^{-B_m^\parallel t} \cos(C_m^\parallel t), \label{eq:correlation_fit_xx}
\end{align}
\end{subequations}
with fitting parameters $A_m^\perp, A_m^\parallel,B_m^\perp, B_m^\parallel,C_m^\perp, C_m^\parallel \in \mathbb R$.
The choice of parameterization in \eqref{eq:correlation_fit_xy} (especially the oscillations' phase) is motivated by the fact that $C^{xy}(0) = 0$, imposed by the spatial symmetries of the spin systems we will consider, that forbid any static correlation between orthogonal current components. Similarly, the choice made in \eqref{eq:correlation_fit_xx} is motivated by the fact that at short times $t \ll (B_m^{\parallel})^{-1}$, the current-current correlation must be maximal at $t=0$ i.e.~$|C^{xx}(t)| \leq |C^{xx}(0)|$.
 The number $M$ of modes in the expansion, $m \in \llbracket 1,M\rrbracket$, must be set by the number of independent peaks observable in the Fourier-transformed averaged currents to ensure all relevant frequencies contributing to the dynamics are captured; it can be chosen in an adaptive way.
 
Because numerical errors accumulate along the integration time, the simulation must be truncated to a finite window $t \in [0,t_{\rm max}]$. To estimate the time $t_{\rm max}$ beyond which the estimated correlations become unreliable, we use the local-in-time requirement $C^{xy}(t) = -C^{yx}(t)$, imposed by broken time reversal and preserved spatial symmetries. This is a physical constraint on the non-equilibrium dynamics that is not strictly enforced by the numerical integration scheme at all steps. Thus, $t_{\rm max}$ may be defined as the point where an estimator of the wrongness of $C^{xy}(t) = -C^{yx}(t)$ becomes too large (see appendix C of Ref.~\cite{mook_spin_2016}).

 Once the correlation function \eqref{eq:Cab} is fitted to the forms \eqref{eq:correlation_fit_xy} and \eqref{eq:correlation_fit_xx} over the time window $t \in [0,t_{\rm max}]$, the Kubo part of the thermal Hall conductivity can be readily obtained from \eqref{eq:kappa_0}:
\begin{equation}
 \kappa_{\rm Kubo}^{xy} = \frac{V}{k_{\rm B}T^2}\int\text dt \,C^{xy}(t)
 \overset{\rm fit}= \frac{V}{k_{\rm B}T^2}\sum_{m=1}^M \frac{A_m^\perp C^\perp_m}{(B_m^\perp)^2 + (C_m^\perp)^2}.
\end{equation}
Similarly for the longitudinal conductivity, 
\begin{equation}
\label{eq/kxx}
 \kappa_{xx} = \frac{V}{k_{\rm B}T^2}\int\text dt \,C^{xx}(t)
 \overset{\rm fit}= \frac{V}{k_{\rm B}T^2}\sum_{m=1}^M \frac{A_m^\parallel B^\parallel_m}{(B_m^\parallel)^2 + (C_m^\perp)^2}. 
\end{equation}

We illustrate this procedure and provide more details in a concrete application case below, see Section \ref{sec:kitaev_afm} and Figure \ref{Fig04}.

\section{Models and results}
\label{sec:results}

We now evaluate the thermal Hall conductivity from classical spin dynamics in two different spin models, depicted in Fig.\ref{fig:square_lattice_dmi}. First, in Sec.~\ref{sec:chiral_fm} we benchmark our methods on a chiral ferromagnetic model that was studied in Ref.~\cite{carnahan_thermal_2021}. Then, in Sec.~\ref{sec:kitaev_afm} we investigate the antiferromagnetic Kitaev model on the honeycomb lattice.

\begin{figure}[b]
  \centering
  \includegraphics[width=\columnwidth]{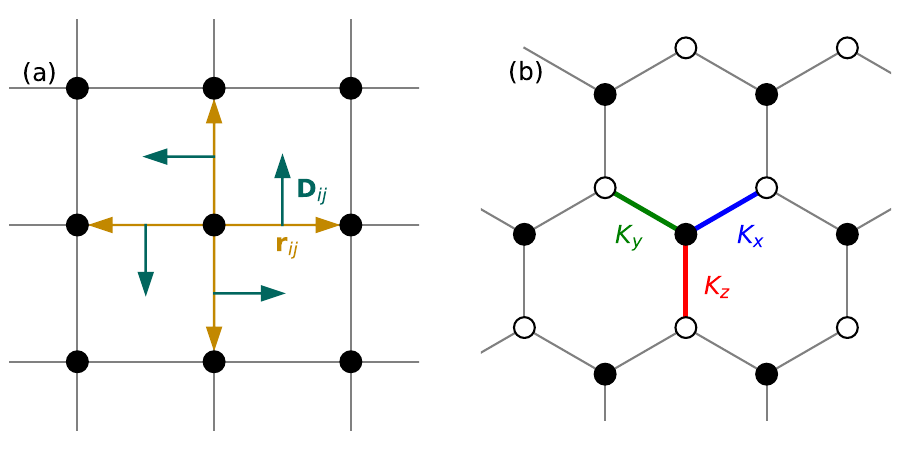}
  \caption{Models studied in this paper and notational conventions. (a) Chiral ferromagnetic model studied in Sec.~\ref{sec:chiral_fm}. The ground state is polarized out of plane and the dark arrows indicate the Dzyaloshinskii-Moriya vectors $\vec{D}_{ij}$ for the direction indicated by the bond vectors $\vec{r}_{ij}$. (b) Antiferromagnetic Kitaev model studied in Sec.~\ref{sec:kitaev_afm}.}
  \label{fig:square_lattice_dmi}
\end{figure}

\subsection{Chiral square-lattice ferromagnet}
\label{sec:chiral_fm}

\subsubsection{Spin model}
\label{sec:phys-mod-A}

We consider the following spin Hamiltonian on the square lattice:
\begin{equation}
\label{eq/spinham}
    \mathcal{H}= J \sum_{\langle i,j \rangle}  \boldsymbol{S}_i \cdot \boldsymbol{S}_j + \sum_{\langle i,j \rangle} \boldsymbol{D}_{ij} \cdot (\boldsymbol{S}_i \times \boldsymbol{S}_j) - \sum_{i} \boldsymbol{B}_{\rm ext} \cdot \boldsymbol{S}_i,
\end{equation}
where $J<0$ is the ferromagnetic Heisenberg exchange, $B_{\rm ext} = B_{\rm ext}\hat z$ is an external magnetic field ($\hat{z}$ the unit vector in the out-of-plane direction), and we assume the Dzyaloshinskii-Moriya interaction vector $\vec{D}_{ij}=D \left( \hat{z} \times \boldsymbol{r}_{ij} \right)$ with $\boldsymbol{r}_{ij}=\bm r_i - \bm r_j$ the vector connecting sites $i$ and $j$ (see panel (a) of Fig.~\ref{fig:square_lattice_dmi}). The interplay of in-plane Dzyaloshinskii–Moriya interactions and a transverse magnetic field is known \cite{han2017skyrmions} to generate chiral magnetic phases, which exhibit spiral textures and skyrmions. Transverse current-current correlations, and consequently thermal Hall conductivity, are generated by thermally excited chiral spin fluctuations, and survive even in the strongly fluctuating (high-temperature) regimes without static spin textures. Thus, any simple description in terms of harmonic modes of magnon-like excitations and the standard formula Eq.\ref{eq:kappaxy} are inadequate.

\subsubsection{Numerical values and implementation}
We use parameter values $J=-1 \, \mathrm{meV}$, $D=0.3 \, \mathrm{meV}$ and $T=0.23 \, \mathrm{K}$ as in Ref.~\cite{carnahan_thermal_2021}, such that the system undergoes a transition from a skyrmion lattice phase at small (but finite) $B$ to a paramagnetic phase at larger $B$. 
We use a time step $\Delta t = 0.657 \, \mathrm{fs}$ (that is to say $\Delta \tau = (J/\hbar) \, \Delta t \approx 0.001$), for both phases of the classical evolution. The results are averaged over 600 independent runs, each one with $n_{\mathrm{eq}} = 6 \cdot 10^7$ time steps for equilibration (i.e. $t_{\rm eq}=n_{\rm eq}\,\Delta t$) and $n_{\mathrm{Kubo}} = 2.4 \cdot 10^8$ steps for the hamiltonian evolution (i.e. $t_{\rm max}=n_{\rm Kubo}\,\Delta t$), recording snapshots of the currents every $1000$ steps. As for extracting the static averages, after a single equilibration we evolve the system for $n_{\mathrm{EM}} = 3 \cdot 10^8$ steps (i.e. $t_{\rm th}=n_{\rm EM}\,\Delta t$) and take snapshots every $n_{\rm snap} = 3000$ steps (i.e. $t_{\rm snap}=n_{\rm snap}\,\Delta t$).

The simulations were performed on systems of size $N = 80 \times 80$ spins. For the Kubo component $\kappa^{xy}_{\rm Kubo}$, we used periodic boundary conditions. For the magnetization component $\kappa^{xy}_{\rm EM}$, we used mixed boundary conditions, periodic along $x$ and open along $y$.

\begin{figure}[t]
    \centering
    \includegraphics[width=\columnwidth]{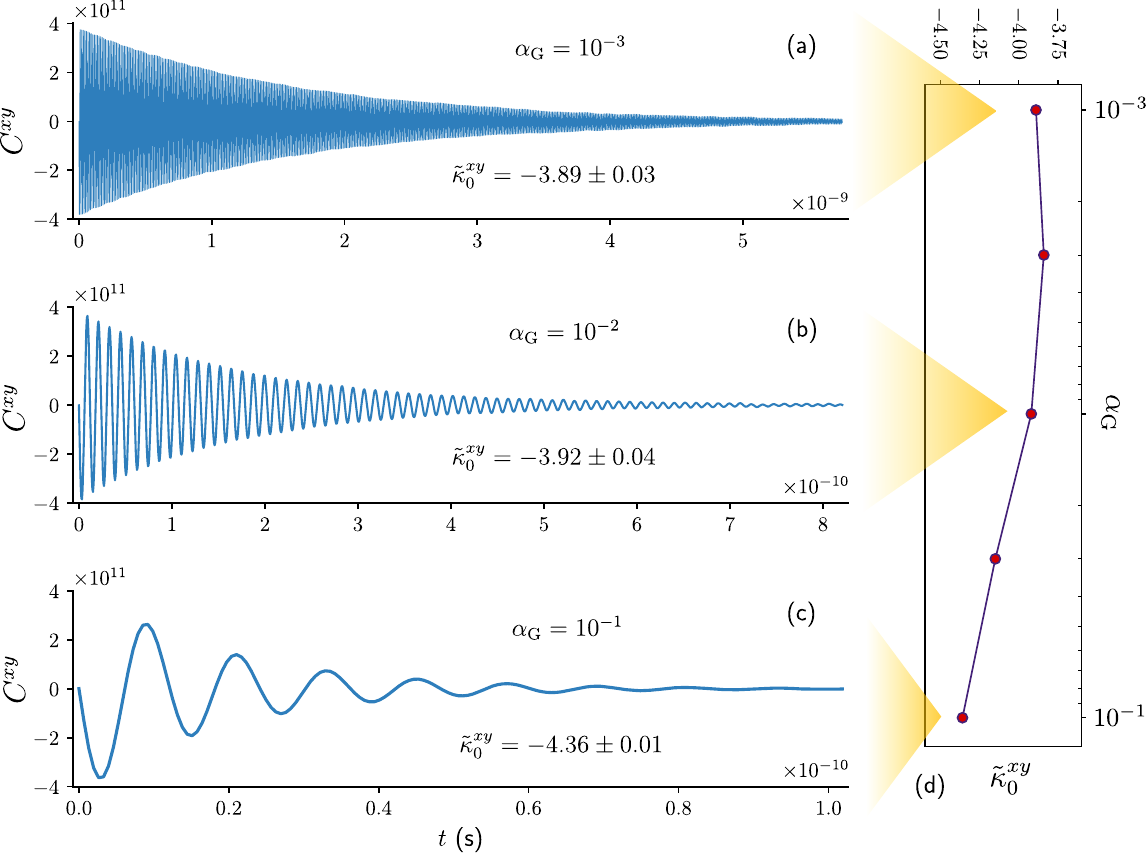}
    \caption{(a-c) Current-current correlation function $C^{xy}(t)$ (in arbitrary units) as a function of time (in s), for different values of the dimensionless Gilbert damping coefficient $\alpha_{\rm G}$ and for $B/J\approx 0.17$. For each curve we indicate the extracted Kubo component $\tilde{\kappa}_0^{xy} = \kappa^{xy} \mu_{\rm B} / \gamma k_{\rm B}$ of the thermal Hall conductivity (in units of $J$). Note the different time scales on the horizontal axes. (d) Kubo component of the Hall conductivity $\tilde{\kappa}_0^{xy}$ as a function of $\alpha_{\mathrm{G}}$, showing deviations from the plateau value for $\alpha_{\mathrm{G}} \gtrsim 10^{-2}$, indicating a breakdown of the nearly-microcanonical regime required for a reliable extraction of the conductivity.}
    \label{fig:alpha_correlations}
\end{figure}

At odds with the physically relevant procedure described in Sec.~\ref{sec:cano-microcano}, we note that Ref.~\cite{carnahan_thermal_2021} maintained a \emph{stochastic} evolution at all times $t \in [-t_{\rm eq},t_{\rm max}]$. 
While this is, in principle, unphysical, we find that it ultimately has little consequence for the estimated physical quantities, while significantly reducing the decay time of current-current correlations $C^{\alpha\beta}(t)$ and thus the required integration time $t_{\rm max}$. We also use this computational trick here, illustrated in Fig.~\ref{fig:alpha_correlations}. At small enough values of the damping parameter $\alpha_{\rm G}$, increasing $\alpha_{\rm G}$ significantly reduces the typical decay time of correlations (thus the computational cost) while the estimated value of $\tilde \kappa^{xy}$ remains identical (within errorbars). At larger values of $\alpha_{\rm G}$, the estimation of $\tilde \kappa^{xy}$ deviates from its correct value, due to numerical instability (e.g. significant non-conservation of energy wihin the Hamiltonian evolution). For these reasons, in the following we use a convenient damping of $\alpha_{\rm G}= 10^{-2}$, having checked that it produces reliable results. However, and crucially, for the Kitaev model investigated in Sec.\ref{sec:kitaev_afm} we \emph{do not} make this assumption (for reasons detailed later) and use instead the physically relevant time evolution procedure, without damping and therefore explicitly enforcing energy conservation, in both phases. 

We note that, although we ultimately use the same system as Ref.~\cite{carnahan_thermal_2021} as a benchmark, our procedure to thermalize the system is different. Indeed, Carnahan et al.\ use classical Monte Carlo sampling in the initialization phase, while we use classical (stochastic) LLG evolution, consistently with the rest of the procedure. In principle, both methods should be perfectly equivalent.

\begin{figure}[b]
    \centering
    \includegraphics[width=\columnwidth]{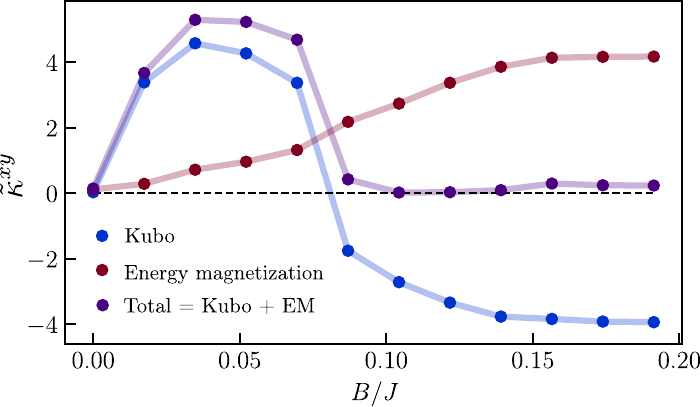}
    \caption{Decomposition of the normalized thermal Hall conductivity $\tilde{\kappa}^{xy} = \kappa^{xy} \mu_{\rm B} / \gamma k_{\rm B}$ of the model Sec.\ref{sec:phys-mod-A} as a function of the magnetic field $B/J$. Shown are the Kubo contribution $\tilde{\kappa}^{xy}_{\rm Kubo}$ (blue), the energy magnetization contribution $\tilde{\kappa}^{xy}_{\rm EM}$ (red), and the total quantity $\tilde{\kappa}^{xy}_{\rm tot} = \tilde{\kappa}^{xy}_{\rm Kubo}+\tilde{\kappa}^{xy}_{\rm EM}$ (purple). Note that the point at $B/J=0$ was computed, not postulated on physical grounds. }
    \label{Fig03}
\end{figure}

\subsubsection{Results}
\label{sec:results_cfm}

Our results for the thermal Hall conductivity are displayed in Fig.~\ref{Fig03}. 
They show excellent agreement with the numerical findings of Ref.~\cite{carnahan_thermal_2021} (see Fig.~3 therein), up to a global factor of 4 that we cannot explain. 
We have carefully checked our analytical derivations and numerical implementations and find no inconsistency on our side, leading us to believe that our results are correct.

At zero field, both contributions $\kappa^{xy}_{\mathrm{Kubo}}$ and $\kappa^{xy}_{\mathrm{EM}}$ vanish. This signals the absence of chiral spin excitations, as a result of preserved time reversal symmetry, not only explicitly but also spontaneously in the low energy states. In particular, there are no static textures with non-zero scalar spin chirality, which otherwise are known to generate magnon chirality and a finite Hall response \cite{PhysRevB.91.125413}. 
Once a finite magnetic field is applied, time-reversal symmetry is explicitly broken and a nonzero Hall response emerges. The total $\kappa^{xy}$ increases with field and reaches a maximum around $B/J \approx 0.05$, which coincides with the regime where skyrmion textures stabilize and chiral spin fluctuations are excited \cite{ezawa2011, banerjee2014, hou2017, carnahan_thermal_2021}.

At higher fields, the Hall response drops towards zero. The transition happens around $B/J \approx 0.08$, as the ground state evolves into a nearly collinear field-polarized state and skyrmion textures disappear. Remarkably, this happens via a rapid change and sign reversal of the Kubo term $\kappa^{xy}_{\rm Kubo}$, whereas the energy magnetization term $\kappa^{xy}_{\rm EM}$ remains smooth and monotonous at the transition. This indicates that the chirality of spin excitations is reflected much more sharply in the current-current \emph{dynamical correlations}, entering $\kappa^{xy}_{\rm Kubo}$, than in the current \emph{static averages}, entering $\kappa^{xy}_{\rm EM}$. Both contributions ultimately saturate to roughly constant and opposite values, as spin fluctuations are progressively gapped out.

We note that in the whole field-polarized region, $\left |\kappa^{xy}_{\rm tot} \right | \ll \left |\kappa^{xy}_{\rm EM} \right |,\left |\kappa^{xy}_{\rm Kubo} \right |  $, because the two contributions are almost equal in absolute value with opposite signs. This makes extracting quantitative information more challenging than in the low-field phase. The model we will consider next (Sec. \ref{sec:kitaev_afm}) shares this feature and will be discussed in detail below.

\subsection{Antiferromagnetic Kitaev model}
\label{sec:kitaev_afm}

\subsubsection{Spin model}
\label{sec:model-B}

Having carefully benchmarked our setup, we now investigate Kitaev's honeycomb model, described by the Hamiltonian
\begin{equation}
\label{eq:kitaev_model}
\mathcal{H} = \sum_{\langle ij \rangle_\gamma} K_\gamma\, S_i^\gamma S_j^\gamma - \sum_i \boldsymbol{B} \cdot \boldsymbol{S}_i,
\end{equation}
where $K_\gamma$ denotes the anisotropic Kitaev couplings along the $\gamma = x, y, z$ bonds of the honeycomb lattice as indicated in Fig.~\ref{fig:square_lattice_dmi} (b), and $\boldsymbol{B}$ is the external magnetic field. 
We consider an antiferromagnetic Kitaev interaction with equal bond strength $K_x=K_y=K_z \equiv J>0$ and a magnetic field pointing along the [111] axis, $\boldsymbol B = B\,(1,1,1)/\sqrt 3$.
Below we briefly remind the reader of the physics of the model and the challenges posed in the range of parameters we wish to investigate.

The dynamics of the nearly field-polarized Kitaev model, at low temperatures, is plausibly well described by spin-wave-like excitations~\cite{mcclarty_topological_magnons_2018,PhysRevB.98.060405}.
The interplay of explicitly broken time-reversal symmetry by the magnetic field, with anisotropic magnetic exchange induced by spin-orbit coupling, induces chirality in the magnon bands as a nonzero Berry curvature. This produces a thermal Hall response even in the absence of nontrivial spin textures \cite{zhang2024thermal, matsumoto_rotational_2011, matsumoto2014,cookmeyer2018, chern2021}, somewhat analogously to the field-polarized phase of the model of Sec.\ref{sec:chiral_fm}. We note that extension of this minimal Kitaev model have been put forward as a possible explanation of the experimentally observed THE in the field polarized phase of $\alpha-$RuCl$_3$~\cite{Czajka2022,chern2021,cookmeyer2018}.  
As temperature increases and larger magnon populations are thermally excited, magnetic polarization is reduced. Besides, close to the saturation field interaction effects become prominent, in particular magnon modes are renormalized \cite{chernyshev2009spin,RevModPhys.85.219} and become less well-defined quasiparticle excitations. These two effects concur to reducing the intrinsic thermal Hall conductivity \cite{koyama2024}, which is thus expected to be a decreasing function of temperature. 
In the language of semiclassical spin dynamics, this means that both the average amplitude of precession motion and the average magnitude of local spin-spin interactions increase with larger thermal fluctuations.

\begin{figure*}[htbp]
    \centering
    \includegraphics[width=\textwidth]{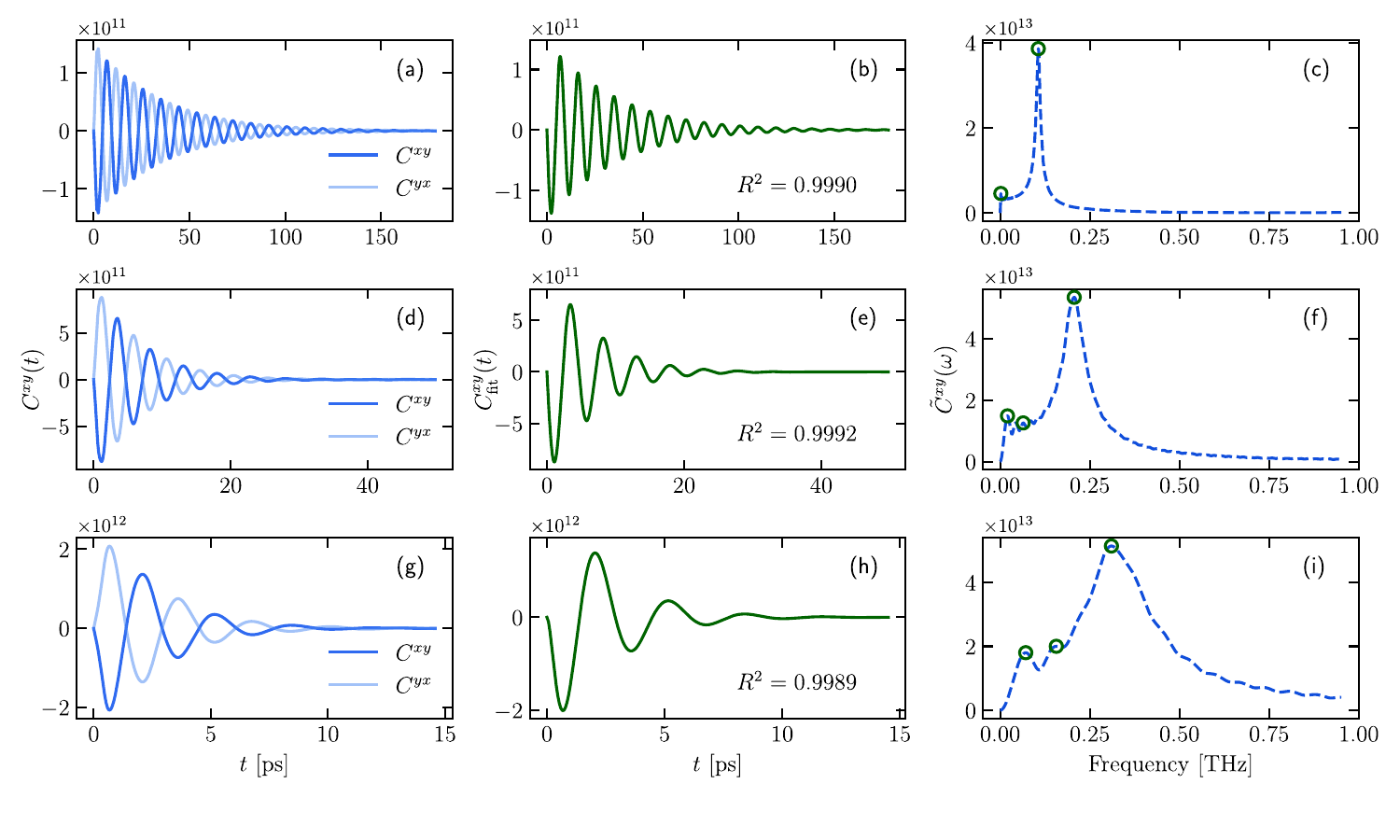}
    \caption{Cross-correlation analysis of the energy currents in the Kitaev model at three representative temperatures: (a–c) $k_{\rm B} T/J \approx 0.026$, (d–f) $k_{\rm B} T/J \approx 0.13$, and (g–i) $k_{\rm B} T/J \approx 0.35$. Left column: raw transverse current correlators $C^{xy}(t)$ and $C^{yx}(t)$. Middle column: fitted correlators $C^{xy}_{\text{fit}}(t)$ using damped sine functions, with $R^2$ values indicating fit quality. Right column: Fourier transforms $\tilde{C}^{xy}(\omega)$, where dominant frequencies are marked by green circles. With increasing temperature, the oscillatory signal is damped more rapidly and spectral peaks broaden, reflecting enhanced magnon loss of coherence due to thermal fluctuations.}
    \label{Fig04}
\end{figure*}

The physics of thermal fluctuations in the model \eqref{eq:kitaev_model} is even more nontrivial below saturation field. In zero field, the anisotropic bond couplings generate an extensively degenerate manifold of classical ground states, highly susceptible to ground state degeneracy lifting by thermal order-by-disorder effects \cite{Baskaran2008}. Remarkably, dynamical spin correlations based on semiclasscial dynamics closely resemble those of the zero temperature quantum solution~\cite{knolle2014dynamics}. Thus, this classical spin liquid regime can be viewed as a precursor to fractionalization in the quantum model~\cite{samarkoon2018}. While a finite field also lifts this extensive degeneracy, even in the intermediate field phases before saturation thermal fluctuations are known to play an important role~\cite{chern2020magnetic}. In particular, nontrivial dynamical modes are excited, that lead to broadened magnon spectra and continuum-like features in the dynamical structure factor \cite{franke2022} beyond simple magnon descriptions.

Here, we investigate this non-integrable regime of finite temperature and intermediate field, with a nearly field-polarized ground state but large  fluctuations and non-linearities. In this regime, all possible excitations, e.g. magnons, are heavily overdamped, possibly losing completely any quasiparticle character.
Therefore, we do not assume \textit{a priori} any kind of quasiparticles, and instead study the semiclassical evolution of the spin system. We note that this method, as an approximation to the spin path integral representation \cite{auerbach2012interacting} that does not explicitly rely on the existence of an ordered ground state nor any mean-field phase, is not \emph{a priori} less well-suited to the challenging problem at hand than other approaches based on the existence of quasiparticles. It also has the advantage of naturally incorporating non-linear interactions between spin degrees of freedom. As such, it may capture essential features of the collective dynamics irrespective of the specific particle basis used to quantize the theory \cite{carnahan_thermal_2021, kim2025}, even in regimes when one or all species of quasiparticles become ill-defined.  This makes the semiclassical method particularly promising in the large field and intermediate temperature regime where its main shortcoming, the wrong classical statistics, is less relevant.

\subsubsection{Numerical values and implementation}

We use $J = 1 \,  \mathrm{meV}$ and set the magnetic field to $B / J= 2.1$, slightly above the critical field necessary for magnon branches to be well-defined \cite{mcclarty_topological_magnons_2018}. 
For the gyromagnetic ratio $\gamma = {\sf g}\mu_{\rm B}/\hbar$ we take ${\sf g}=2$ the electronic g-factor. We use a time step $\Delta t = 0.657 \, \mathrm{fs}$ (that is to say $\Delta \tau = (J/\hbar) \, \Delta t \approx 0.001$) for both phases of the classical evolution. Depending on $T$, the results are averaged over a variable number (ranging from 600 to 3000) of independent runs, with $n_{\mathrm{eq}} = 6 \cdot 10^7$ time steps for equilibration (i.e. $t_{\rm eq}=n_{\rm eq}\,\Delta t$), with damping $\alpha_{\rm G} = 0.01$ only in the equilibration phase, and $n_{\mathrm{Kubo}} = 2.4 \cdot 10^7$ steps for the hamiltonian evolution (i.e. $t_{\rm max}=n_{\rm Kubo}\,\Delta t$), recording snapshots of the currents every $100$ steps. As for extracting the static averages, after a single equilibration we evolve the system for $n_{\mathrm{EM}} = 1 \cdot 10^9$ steps (i.e. $t_{\rm th}=n_{\rm EM}\,\Delta t$) and take snapshots every $n_{\rm snap} = 4000$ steps (i.e. $t_{\rm snap}=n_{\rm snap}\,\Delta t$).

The simulations were performed on systems of size $N = 80 \times 80$ unit cells (two spins per unit cell). For the Kubo component $\kappa^{xy}_{\rm Kubo}$ and the momentum-space definition of the energy magnetization contribution in \eqref{eq:kappa_1m}, we used periodic boundary conditions. For the real-space definition of the energy magnetization contribution in \eqref{eq:kappa_1}, we used mixed boundary conditions, periodic along $x$ and open along $y$.

As a final note, because the spin system we simulate is two-dimensional, all conductivities are plotted in units of W/K, where $\kappa^{xy} \sim 10^{-12}$ W/K. To obtain the physical quantity for a three-dimensional crystal, one must divide by the interlayer distance, which for $\alpha$-RuCl$_3$ would be ${\sf a} = 5.7$ \AA, corresponding to $\kappa^{xy} \sim  10^{-3}-10^{-2}$ W/K/m.

\begin{figure}[t]
    \centering
    \includegraphics[width=\columnwidth]{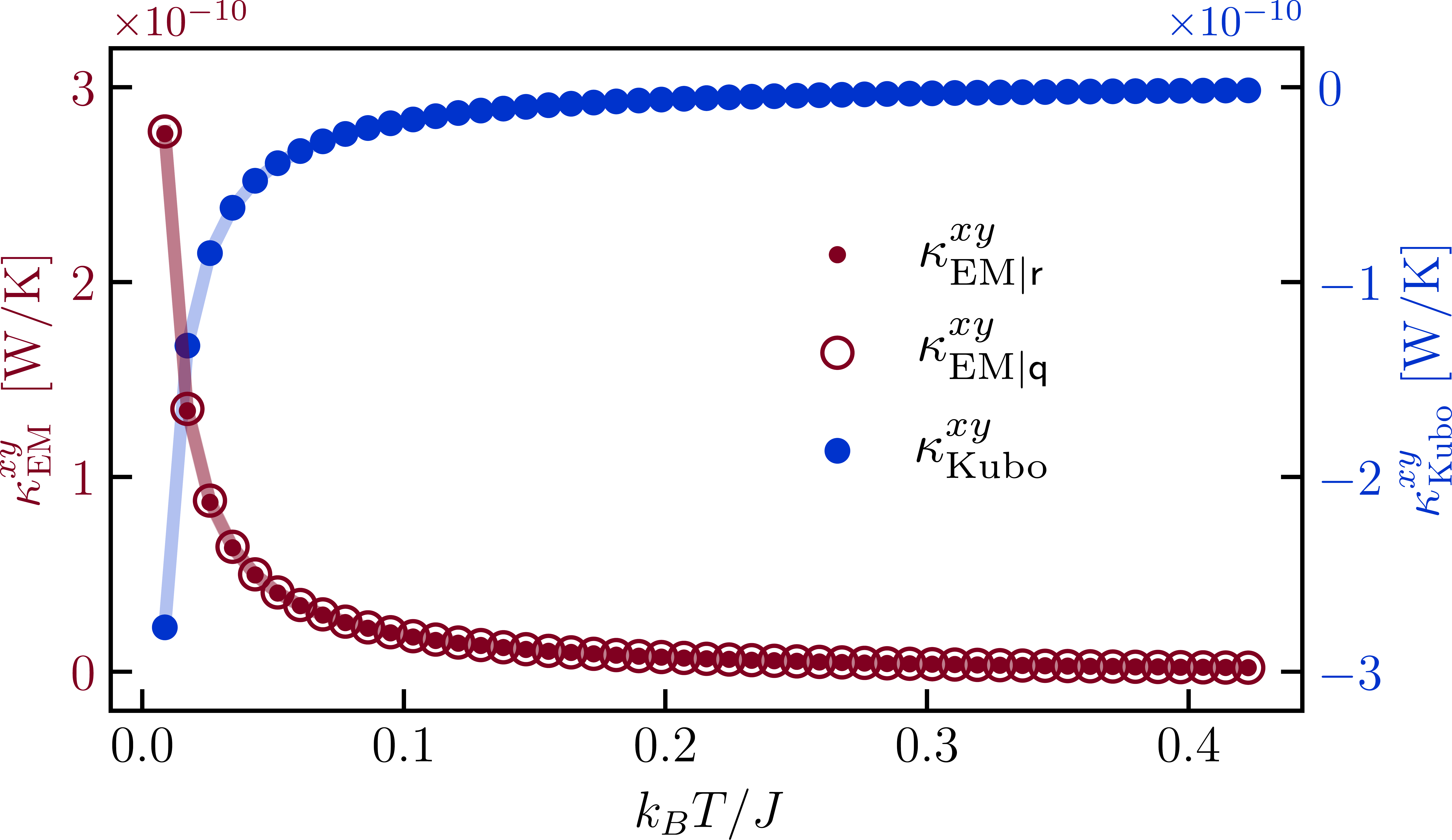}
    \caption{
    Temperature dependence of the thermal Hall conductivity components in the model Sec.\ref{sec:model-B} at $B/J$ = 2.1. Blue (axis on the right): the Kubo contribution ${\kappa}^{xy}_{\mathrm{Kubo}}$, discussed in Sec.\ref{sec:kit-kubo}. Red (axis on the left): the energy magnetization contribution ${\kappa}^{xy}_{\mathrm{EM}}$, evaluated from the real-space and the momentum-space methods (filled and empty markers, respectively), discussed in Sec.\ref{sec:kit-em}. Errorbars would generally be smaller than the marker size and are not displayed.}
    \label{Fig05}
\end{figure}

\subsubsection{Kubo contribution}
\label{sec:kit-kubo}

To understand the emergence of $\kappa_{\rm Kubo}^{xy}$ in the thermal Hall response, we now analyze the cross-correlation functions of the energy currents, defined in \eqref{eq:Cab}, at three different temperatures, $k_{\rm B} T / J \approx 0.026, \, 0.13 \, \text{and} \, 0.35$. They are shown in Fig. \ref{Fig04}. Panels (a), (d), (g) display the raw data for the correlation functions  $C^{xy}(t)$ from $t=0$ to $t=t_{\rm max}$, their respective cutoff time determined by the procedure described in Sec.\ref{sec:fitting}. We note that $C^{yx}(t)$ displays the expected antisymmetric relation, $C^{xy}(t) \simeq -C^{yx}(t)$, as a consequence of symmetry in the system's dynamics -- and up to numerical errors that accumulate at later times (see discussion in Sec. \ref{sec:fitting}).

From these raw cross-correlations, we then apply the fitting procedure described in section \ref{sec:fitting}. As explained around \eqref{eq:correlation_fit_xy}, $C^{xy}(t)$ is approximated as a sum of damped sine functions, by identifying dominant frequencies in the Fourier transform of the raw signals, as illustrated in panels (c), (f) and (i). We perform this analysis, even though the raw correlation functions are smooth and well-behaved, so that a direct numerical integration would be possible in principle. This allows us to (1) enforce exact constraints such as $C^{xy}(0)=0$, and (2) obtain controlled error estimates by bootstrapping methods.
The fits, shown in panels (b), (e), and (h) of Fig. \ref{Fig04}, reproduce the raw correlators with high accuracy ($R^2 \gtrsim 0.99$). 

The presence of long-lived oscillations in the transverse current correlation function $ C^{xy}(t)$ indicates the existence of coherent spin excitations, that at low temperatures can be identified as a sharp magnon mode. This appears as a well-defined sharp peak in the frequency domain (right-column of Fig. \ref{Fig04}, especially first row), marked with a green circle. Other minor peaks, also marked, appear in the analysis, although their physical interpretation is not as straightforward.

As $T$ increases, the oscillatory signal $C^{xy}(t)$ decays faster (see scales on the horizontal axes, left-column of Fig. \ref{Fig04}, from top to bottom). This corresponds in the frequency domain to a broadening of the Lorentzian peak, and corresponds physically to the magnon mode becoming more and more ill-defined as the imaginary part of its self-energy $\Sigma(\omega)$ becomes large, and ultimately ${\rm Im}\Sigma(\omega)/\omega = O(1)$. 
The lorentzian peak also undergoes a significant shift as $T$ increases, which corresponds to a significant real part of the self-energy, ultimately ${\rm Re}\Sigma(\omega)/\omega = O(1)$. Both facts signal a significant renormalization of the excitation spectrum and magnon decoherence, that can be understood as a consequence of strong magnon-magnon interactions at high enough $T$, which we argued is well captured by semiclassical spin dynamics.

Integrating the fitted functions $C^{xy}_{\text{fit}}(t)$ according to \eqref{eq:kappa_0} yields the Kubo contribution to the Hall conductivity: see Fig.~\ref{Fig05}. 
In practice, obtaining a well-converged integral requires simulation times long enough to capture the decaying tail of the signal, followed by extensive sample averaging (using up to 3000 samples at the lowest temperature). At lower temperatures, where currents and their correlations decay much more slowly, long simulation times and subsequently larger statistical variance concur in increasing the size of error bars at low $T$.

\subsubsection{Energy magnetization contribution}
\label{sec:kit-em}

In Fig.\ref{Fig05} we show the evaluation of the energy magnetization contribution $\kappa^{xy}_{\rm EM}$ from both the real-space definition [given by \eqref{eq:kappa_1}] and the momentum-space definition [given in \eqref{eq:kappa_1m}]. 
Notably, both the Kubo contribution $\kappa^{xy}_{\rm Kubo}$ and the magnetization contribution $\kappa^{xy}_{\rm EM}$ diverge as $1/T$ in the low-temperature limit, meanwhile their sum (the physical quantity) remains finite -- see the discussion in Sec.~\ref{sec:lrt_hall}. 
This is also what happens in the model we studied in Sec.\ref{sec:chiral_fm} (see e.g. Fig. 2a in Ref.\cite{carnahan_thermal_2021}), and is required on physical grounds.

Besides, the two different methods of evaluation \eqref{eq:kappa_1} and \eqref{eq:kappa_1m} yield two different quantities $\kappa^{xy}_{\rm EM|\sf r}$ and $\kappa^{xy}_{\rm EM|\sf q}$. We can check that their difference remains zero within error bars, see Fig. \ref{Fig06} and the discussion below. This confirms the robustness of our numerical evaluations -- indeed, this makes $\kappa^{xy}$ independent of the method of numerical extraction, which is of course necessary in theory but non-trivial in practice.

In earlier work reported in Ref.~\cite{Kim2024}, the real-space method was found unsuitable for triangular lattice geometries, with the result being dependent on the lattice configuration (see Supplemental Material of Ref.~\cite{Kim2024}). We have \emph{not} encountered the same problem for the honeycomb model and find that the results of both methods agree within error bars, for sufficiently large and converged systems. 

A typical equilibrium spatial profile of the energy magnetization currents at low temperatures is shown in Fig. \ref{profile}. The chain-resolved current is defined as
\begin{align}
    \label{eq:chainresolved}
   j^x_l=  \sum_{i \in l}\sum_{j<i}  j^{\perp}_{j\rightarrow i},
\end{align}
where the summation $i \in l$ is over all sites $i$ belonging to the chain with index $l$, the summation $j<i$ is over sites $j$ from \emph{all} chains (but in practice only the same chain $l$ and neighboring chains will contribute),
and $\perp$ indicates the component of currents in the direction perpendicular to the $z$ bonds (and thus parallel to the edge). The energy magnetization current is strongly localized near the system boundaries, while it only consists of very small thermal noise in the bulk.

\begin{figure}[t]
    \centering
    \includegraphics[width=\columnwidth]{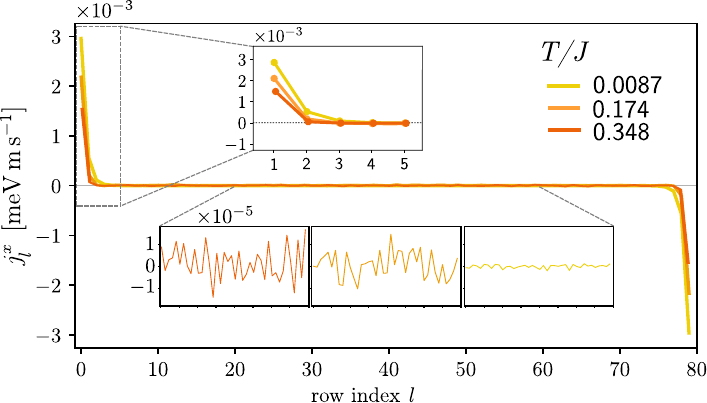}
    \caption{Chain-resolved energy magnetization current $j_l^x$ as a function of the unit-cell chain index $l$ at three different temperatures. The current is strongly localized near the edges of the sample, indicative of gapped excitations. In the bulk, currents are very small and fluctuating, and are essentially thermal noise (vanishing at small temperatures).}
    \label{profile}
\end{figure}

\subsubsection{Total thermal Hall conductivity}
\label{sec/total-thc}

We now discuss the total thermal Hall conductivity $\kappa^{xy}(T)=\kappa^{xy}_{\rm Kubo}+\kappa^{xy}_{\rm EM}$, displayed in Fig. \ref{Fig06}. 
In the low-temperature regime, sharp spin-wave modes as chiral energy carriers are thermally excited, therefore $\kappa^{xy}(T)$ is expected to increase in this regime. Here we find within errorbars a peak value for $k_{\rm B} T/J \lesssim 0.05$, which we attribute to an enhanced activation by classical statistics, see discussion below and in Sec.\ref{sec:comparison-lswt}. 
At higher temperatures, an increased effect of interactions and of anharmonicities, the loss of well-defined quasiparticles, and the reduction of the average magnetic moment along the field all contribute to reducing $\kappa^{xy}(T)$, ultimately suppressing the Hall response completely in the limit $T \rightarrow \infty$. 

We note that errorbars, which even after heavy sample averaging (3000 samples for the lowest temperature and around 600 for larger $T$) are of the order of $0.5 - 1\%$ of the estimated integrals, become significant in the very small-$T$ limit as both components $\kappa^{xy}_{\rm Kubo}$ and $\kappa^{xy}_{\rm EM}$ become large. Because the net conductivity $\kappa^{xy}$ remains finite, errorbars become proportionally large in that $T \rightarrow 0$ limit, as can be seen on Fig. \ref{Fig06}. That they also become larger in absolute values in that limit was discussed in Sec.\ref{sec:kit-kubo}.

\begin{figure}[t]
    \centering
    \includegraphics[width=\columnwidth]{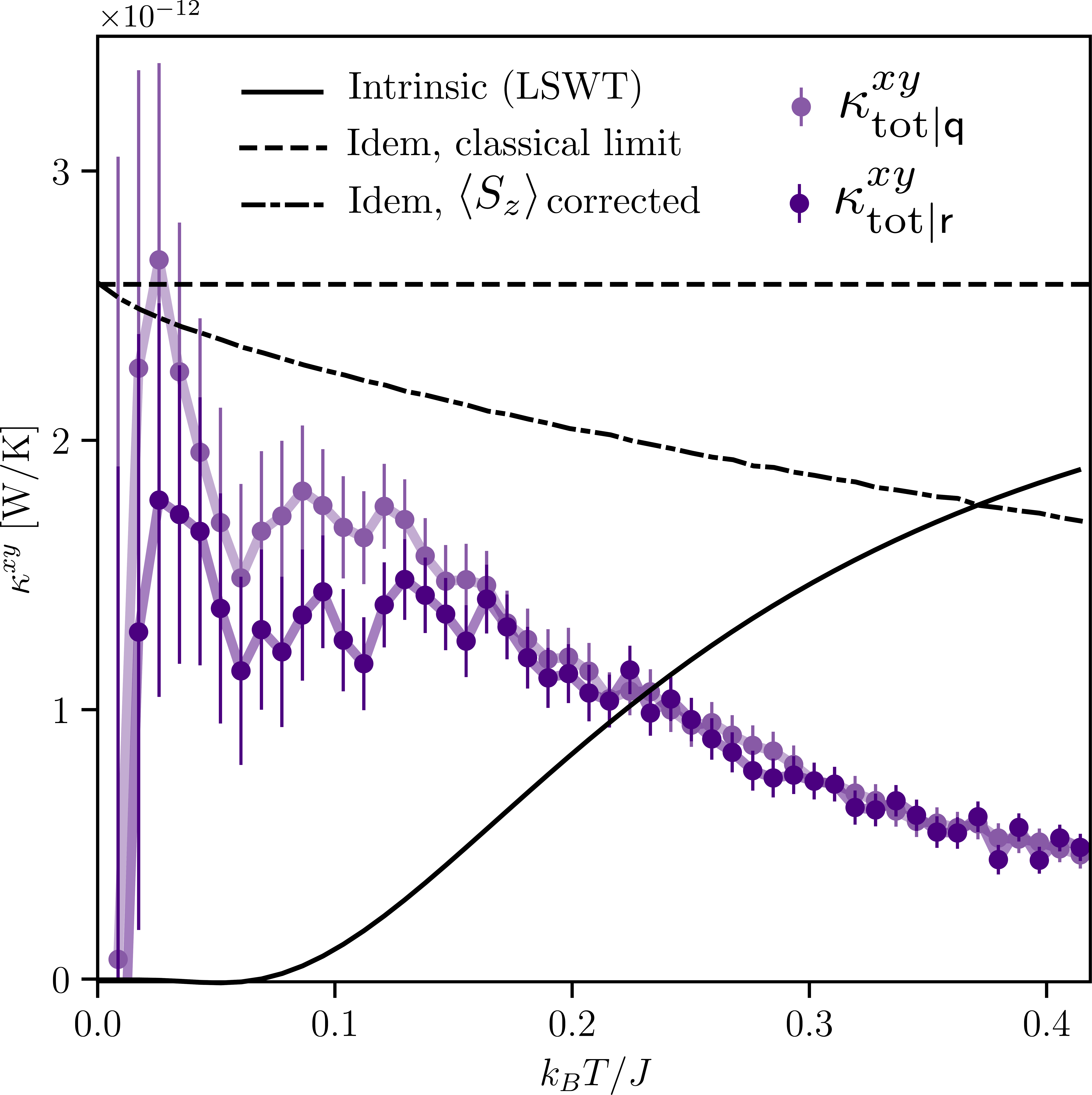}
    \caption{Total thermal Hall conductivity ${\kappa}^{xy}_{\rm tot}$ of the model Sec.\ref{sec:model-B}, obtained from our semiclassical spin dynamics simulations, both with $\sf (r)$ the real space and $(\sf q)$ the momentum space definition of the energy magnetization contribution: see discussion in Sec.\ref{sec/total-thc}. For comparison we plot the intrinsic magnon thermal Hall conductivity obtained from linear spin-wave theory, and two variations thereof obtained by considering a classical statistics for excitations and further correction of the magnetization along $z=[111]$: see discussion in Sec.\ref{sec:comparison-lswt}.}
    \label{Fig06}
\end{figure}

The behavior at low temperatures, close to and below the magnon gap $\Delta_{\rm m} = 0.1J$, should be taken cautiously. Although we indeed find that $\kappa_{xy}$ vanishes within errorbars in the $T\rightarrow 0$ limit, the magnitude of $\kappa_{xy}$ in that low-$T$ region is surprising, and might be caused by the classical statistics of excitations imposed by our semiclassical stochastic method. Indeed, classical statistics is known \cite{bergqvist2018} to overestimate particle densities in the low-$T$ regime. Meanwhile, the rest of the curve, for $k_{\rm B} T \gtrsim 0.1J$, should not be seriously affected by the difference between classical and bosonic statistics. In particular, this is definitely not what explains the discrepancy with the intrinsic free-magnon curve: for a discussion of that point, see Sec.\ref{sec:comparison-lswt}.

\subsubsection{Longitudinal thermal conductivity}
\label{sec:kxx}

\begin{figure}[t]
    \centering
    \includegraphics[width=\columnwidth]{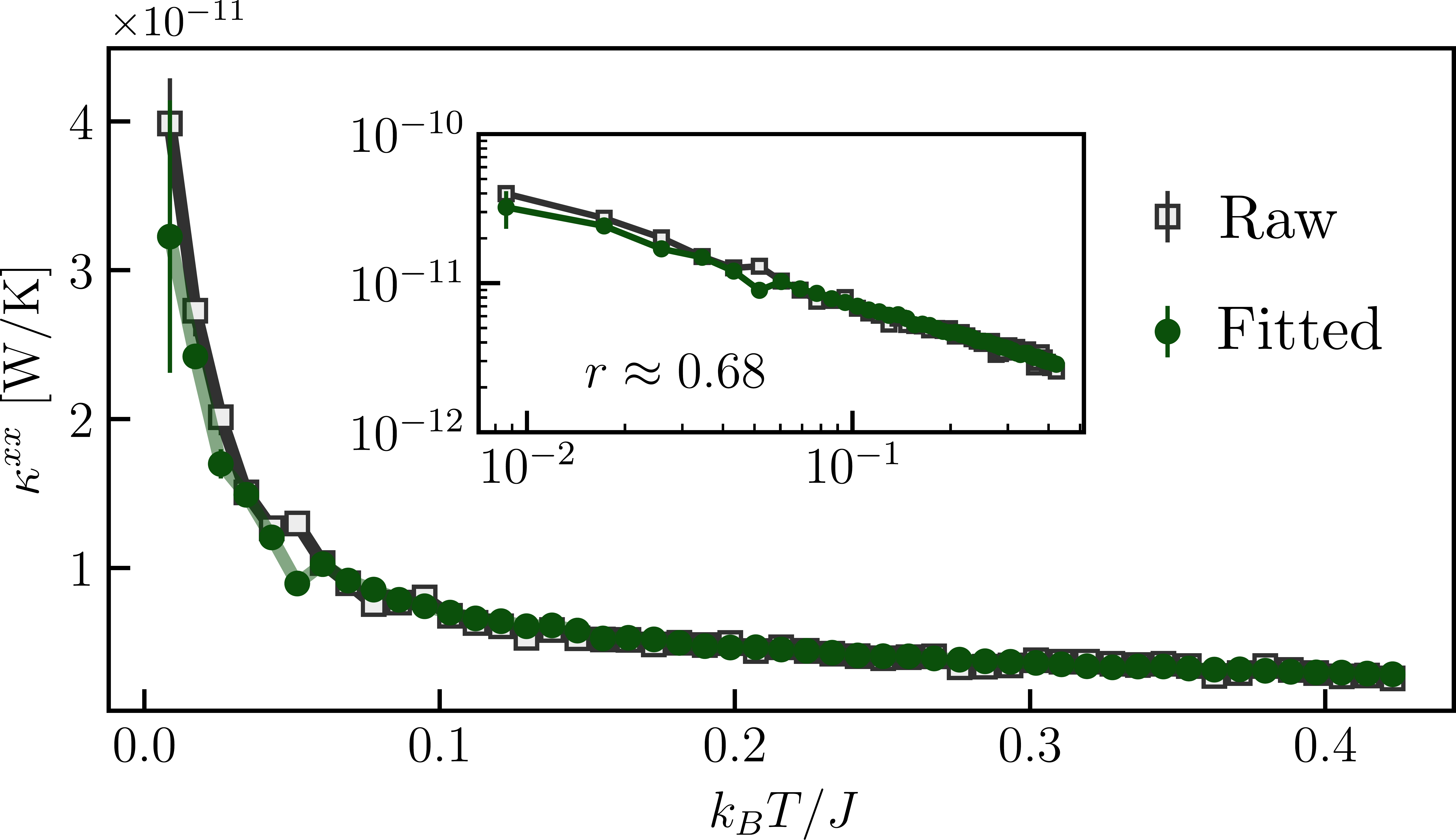}
    \caption{ Longitudinal thermal conductivity ${\kappa}^{xx}$ of the model \eqref{eq:kitaev_model} for $B/J = 2.1$ as a function of temperature, (main) linear scale, (inset) log-log scale. Conductivities extracted by integrating both raw current-current correlations $C^{xx}(t)$ and fitted data $C^{xx}_{\rm fit}(t)$ are shown.}
    \label{Fig07}
\end{figure}

From the current-current correlations discussed in Sec.\ref{sec:kit-kubo}, we also extract the longitudinal thermal conductivity $\kappa_{xx}$ from \eqref{eq/kxx}, both from the fitting procedure described there and from directly computing the time integral $\int\text dt \,C^{xx}(t)$ without any fitting. Generally, we find that the fitting procedure yields much more stable results (i.e.\ less dependent on the number of terms included in the sum of Eqs.\ \ref{eqs:28}) for the cross-correlation $\tilde C_{xy}$ than for $\tilde C_{xx}$. Thus, for computing $\kappa_{xx}$ we keep both methods as a cross-check: the results are displayed in Fig. \ref{Fig07}. 

In a wide temperature range, we find $\kappa_{xx} \sim 1/T^r$ with $r \approx 0.68$. There does not seem to be a straightforward understanding of this in terms of $\kappa \propto c_{v} v^2 \tau_{\rm rel}$ \cite{tritt2005thermal} with $c_v$ the carriers' specific heat, $v^2$ their mean square velocity and $\tau_{\rm rel}$ the current's relaxation time. At much higher temperatures $T \gtrsim J$, one expects a crossover to $\kappa \propto 1/T$, since there the only relevant inverse time scale becomes $\tau_{\rm rel}^{-1}\sim T$. Meanwhile, we find another power-law behavior which persists down to much lower temperatures, even below the magnon gap $\Delta_{\rm m} = 0.1 J$. Indeed, as our simulated spin model \eqref{eq/spinham} does not include any extrinsic source of scattering, the relaxation of energy currents can only originate from umklapp (magnon-magnon) scattering \cite{ziman2001electrons} and the scaling law $\kappa_{xx} \sim 1/T^r$ holds even in the $T \rightarrow 0$ limit. This could indicate that a unique mechanism is responsible for thermal transport in the whole range of temperatures investigated. 

We note that in our spin only model, the ratio $\kappa_{xy}/\kappa_{xx}$ is as large as $O(1)$. This can be understood as a consequence of the longitudinal resistivity $\rho_{xx}$ being small, so that diagonal and transverse elements of $\boldsymbol \sigma = \boldsymbol \rho^{-1}$ can be of comparable size.
Of course, our numerical model does not include phonons, which are generally the main energy carriers in insulators: taking phonons into account would considerably increase the value of $\kappa_{xx}$.

\subsection{Discussion}
\label{sec/discussion}

In the following, we consider our results in a broader perspective. Using Kitaev's honeycomb model \eqref{eq:kitaev_model} as a case study, we discuss what aspects of thermal Hall transport can (or cannot) generally be captured by semiclassical spin dynamics. We focus on two points: (1) the comparison with the pure intrinsic contribution within LSWT and the role of effects beyond the harmonic approximation resulting in Eq.\ref{eq:kappaxy}; and (2) the (in)validity of an alternative method to extract $\kappa^{xy}$ purely from static averages of equilibrium currents~\cite{tang2019quantized,okubo2025thermal}. 

\subsubsection{Comparison to the intrinsic THE within linear spin-wave theory}
\label{sec:comparison-lswt}

For comparison, in Fig.\ref{Fig08} we also show the intrinsic thermal Hall conductivity predicted within linear spin-wave theory. We used the same spin-wave expansion as in Ref.\cite{mcclarty_topological_magnons_2018}, which yields magnon bands with a Berry curvature and an intrinsic thermal Hall conductivity, see Sec.\ref{sec:intrinsic-lsw} and \eqref{eq:kappaxy} therein. In this formula, temperature enters solely through the thermal distribution of magnons, without consideration of any many-body effects. As a result, \eqref{eq:kappaxy} predicts a monotonic increase of $\kappa^{xy}$ with temperature, as higher-energy modes become populated and contribute additively to the transverse thermal transport. This is at odds with our numerical results, where $\kappa_{xy}(T)$ displays a non-monotonic and mostly \emph{decreasing} behavior, with a maximum value at low temperatures followed by a roughly $1/T$ decrease. What explains this discrepancy?

First we note that remarkably, both the intrinsic LSWT formula and our numerical evaluation yield the same order of magnitude for $\kappa^{xy}$ at intermediate temperatures, and more strikingly the peak value we obtain at low temperatures equals the $T\rightarrow \infty$ limit of the intrinsic LSWT result. We understand this as a consequence of our semiclassical spin dynamics effectively imposing classical (Rayleigh-Jeans) statistics $n_{\rm cl}(\omega) = k_{\rm B} T / \omega$ for the spectrum of excitations, because of the numerical implementation of thermal noise as described in Sec.\ref{sec:spin_dyn}. Because this overpopulates low-energy modes compared to Bose statistics, it produces a finite $\kappa^{xy}$ already at very low $T$, where it is consistent with the classical version of the intrinsic $\kappa^{xy}$ formula (case $c_2(\rho)=1/\rho$, dashes in Fig.~\ref{Fig06}). This difference of statistics explains the absence of activation behavior at low $T$. By contrast, the intrinsic LSWT formula with Bose statistics (solid curve in Fig.~\ref{Fig06}) predicts thermal activation governed by the magnon gap $\Delta_{\rm m}$ above which the Berry curvature is concentrated.

Having understood the role of statistics from the low-$T$ behavior, we now consider the high-$T$ regime. There as well, our numerical results disagree both with the intrinsic LSWT prediction and its counterpart for classical statistics. For reference, we compare our numerical results not only to the classical limit of LSWT, but also to the same classical expression multiplied by the reduced magnetization $\langle S_z\rangle_T$, as a proxy for the effect of magnetization depletion within a non-interacting magnon picture (dot-dashed curve in Fig.~\ref{Fig06}). In this scenario, the only temperature dependence of $\kappa^{xy}$ stems from the reduction of the ordered moment by thermal fluctuations, while the Berry curvature and quasiparticle weight are left unchanged. While this correction captures part of the overall downward trend found numerically, it still significantly overestimates $\kappa^{xy}(T)$ at intermediate and high temperatures. This indicates that the suppression of the thermal Hall conductivity cannot be attributed exclusively to the reduction of magnetization. 

Instead, we explain this discrepancy from shortcomings of the intrinsic free-particle picture as temperature increases. Indeed, in the high temperature regime, spin fluctuations are strong and the non-interacting magnon approximation is no longer valid (see especially discussion in Sec.\ref{sec:kit-kubo}). Single-particle modes are strongly renormalized and broadened, and lose overall spectral weight towards the incoherent continuum, which may result in their decreased efficiency at carrying energy, and in the suppression of $\kappa^{xy}$ at larger $T$, as we indeed find numerically. Additionally, non-intrinsic many-body mechanisms such as skew-scattering contributions \cite{mangeolle2022phonon,mangeolle2022thermal,dimos2025thermal} also contribute to the THE (from the thermal Hall \emph{resistivity} $\rho_{xy}$). Because $\kappa_{xy}\approx -\rho_{xy}\kappa_{xx}^2$ and we find a large $\kappa_{xx}$ at low $T$ (see Sec.\ref{sec:kxx} and Fig.\ref{Fig07} therein), these effects can be non-negligible even at quite low temperatures. Overall, because semiclassical spin dynamics described by the LLG equation incorporates non-linear interaction effects of spin wave excitations, it automatically captures these effects. 
Thus, we argue it may more faithfully capture the physics of thermal Hall conductivity in frustrated spin systems, especially at high temperatures, than \eqref{eq:kappaxy} -- provided the physical statistics of excitations is faithfully reproduced, which we discuss in the conclusion.

\subsubsection{Hall conductivity extracted from equilibrium (magnetization) edge currents}

During the preparation of this work, we became aware of Ref.~\cite{okubo2025thermal}, which (similar to Refs.~\cite{vinkler2018approximately,tang2019quantized,PhysRevB.107.L220406} for a fermionic system) proposes to compute the thermal Hall conductivity from an \emph{equilibrium} evaluation of the edge energy currents. Using the same geometry as their Fig. 1b, with zig-zag chains of alternating $x$ and $y$ bonds labeled by $l \in \llbracket 0, N_z \rrbracket$ and separated by rows of $z$ bonds, we define the equilibrium current in the upper ($>$) and lower ($<$) halves of the system as
\begin{align}
\label{eq:Jedge}
I_{\rm eq,>}^{\perp} = \frac 2 V \sum_{l =0}^{N_z/2} j^x_l, \quad
 I_{\rm eq,<}^{\perp} = \frac 2 V \sum_{l =N_z/2}^{N_z} j^x_l, 
\end{align}
where $N_z$ is the total number of zig-zag chains, and we recall the definition of chain-resolved currents \eqref{eq:chainresolved}.

The antisymmetrized edge current, at a given \emph{uniform} temperature $T$, is then defined as 
\begin{equation}
    I_{\rm eq} = (I_{\rm eq,>}^{\perp} - I_{\rm eq,<}^{\perp})/2.
\end{equation}
Computing the derivative of this current with respect to temperature $T$, one may extract a contribution to the thermal Hall conductivity as
\begin{equation}
\label{eq:kedge}
\kappa_{\rm eq}^{xy} = L_z \partial_T I_{\rm eq},
\end{equation}
where $L_z$ is the width of the system in the direction parallel to the $z$ bonds. The result is plotted in Fig.\ref{Fig08}.
We also computed the quantity $\kappa_{\rm edge}^{xy}$ defined similarly but keeping only currents flowing purely along the edge, i.e.\ only terms at $l=0$ and $l=N_z$ in the summation of Eq.\ref{eq:Jedge}. We find that this yields virtually identical results, which means that magnetization currents flow almost exclusively along the edges. There is no theoretical justification \textit{a priori} for this equality $\kappa_{\rm edge}^{xy}\simeq \kappa_{\rm eq}^{xy}$, which we report here as a numerical observation.

Because this approach relies on evaluating equilibrium currents, neglecting the linear response term $\kappa_{\rm Kubo}^{xy}$ altogether, it corresponds physically to computing only the energy magnetization current. This is what we illustrate in Fig.\ref{Fig08}, where we show that both contributions can actually be identified, $\kappa_{\rm eq}^{xy} \simeq \kappa_{\rm EM}^{xy}$ -- as a numerical fact backed by theoretical intuition. Therefore, computing $\kappa_{\rm eq}^{xy}$ (or $\kappa_{\rm edge}^{xy}$) alone does not give access to the thermal Hall conductivity $\kappa_{\rm tot}^{xy}$, which is the only physically relevant (measurable) transport quantity. In particular, the $1/T$ divergence of $\kappa_{\rm eq}^{xy}$ at low temperatures, also witnessed in the classical Monte Carlo evaluation of Ref.~\cite{okubo2025thermal} (see their Fig. 16c), is unphysical and signals that a compensating contribution to $\kappa^{xy}$ is missing -- namely, the linear response term $\kappa_{\rm Kubo}^{xy}$.

We note that in gapped fermionic systems, where transport properties are dominated by edge states and the bulk is gapped, an equilibrium edge-current approach can be justified (see Refs.~\cite{tang2019quantized,okubo2025thermal,PhysRevB.107.L220406,vinkler2018approximately}). In all other cases, including gapped or gapless bosonic systems, this approach generally fails, as we illustrated here.

\begin{figure}[t]
    \centering
    \includegraphics[width=\columnwidth]{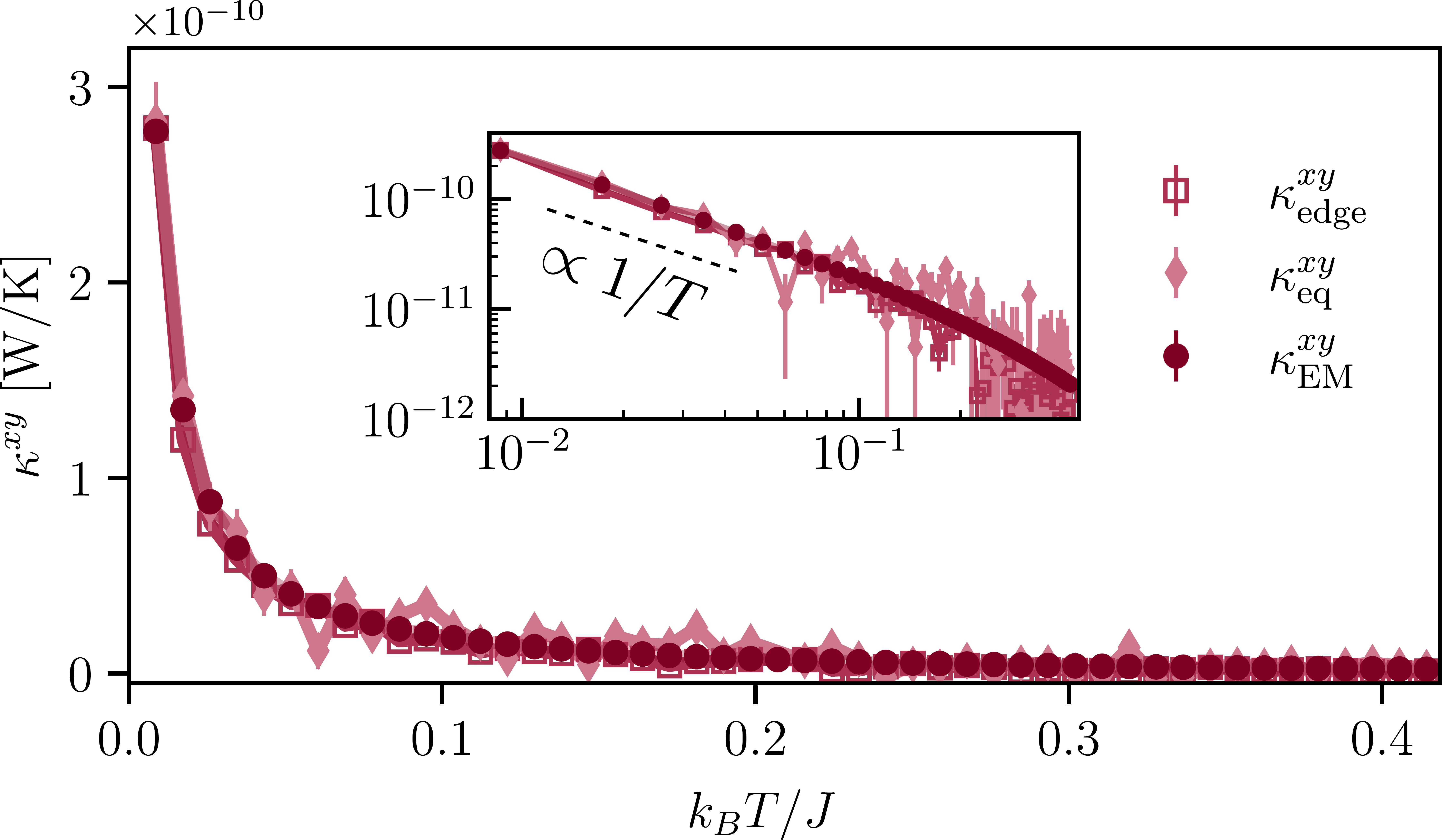}
    \caption{Contribution $\kappa_{\rm eq}^{xy}$ to the thermal Hall conductivity, computed with the method of Ref.~\cite{okubo2025thermal} using \eqref{eq:kedge}, compared with the sole contribution from the edge sites $\kappa_{\rm edge}^{xy}$, and with the energy magnetization contribution $\kappa_{\rm EM}^{xy}$ reproduced from Fig.\ref{Fig05}.
Inset: log-log scale with $1/T$ slope.}
    \label{Fig08}
\end{figure}

\section{Conclusion}
\label{sec/concl}

We have presented a semiclassical method to calculating the thermal Hall conductivity in spin systems, based on Landau–Lifshitz–Gilbert dynamics, for general lattice geometries and interaction types. Our approach is based on the expression of local energy currents purely in terms of spin operators, see \eqref{eq/poissonbracket}. 
The thermal currents are generally the sum of two components \eqref{eq:kappa_tot}, an equilibrium (magnetization) term and a linear-response (Kubo) term. The latter involves dynamical current-current correlators, the proper evaluation of which requires heavy sample averaging and large system sizes, such as can be reached thanks to semiclassical time evolution. We applied this method to the antiferromagnetic Kitaev honeycomb model in a magnetic field. Following what appears as a general rule, we find that both the magnetization and the Kubo contribution to the thermal Hall conductivity diverge as $1/T$ at low temperatures, while their sum remains finite. We developed a systematic numerical procedure to ensure the cancellation between these two divergences, and thus extract the relevant physical quantities. Our procedure extends well beyond the model we considered, and allows to compute the thermal Hall conductivity of any polarized spin system in the low-temperature limit.

 Our analysis shows that in Kitaev's AF model above the polarization field, the thermal Hall conductivity originates from coherent spin-wave excitations at low temperatures, which give rise to well-defined frequency-resolved current-current correlations. However, our numerical results deviate from the prediction for the purely intrinsic contribution obtained from non-interacting magnons at all temperatures, although the overall magnitude remains comparable. At low temperatures, this discrepancy instead originates from the implementation of thermal equilibrium in our classical simulations: the imposed classical (instead of bosonic) statistics overpopulates low-energy modes and leads to a finite $\kappa^{xy}$ even at temperatures below the magnon gap. At higher temperatures, we find that the deviation from the free-magnon prediction cannot be explained by a mere difference of statistics, not even when including corrections to the size of magnetic moments by fluctuations. We find instead that our numerical results display a significant effect of interactions beyond the harmonic approximation, including magnon–magnon scattering and the gradual loss of coherent quasiparticles. Overall, the differences between our numerical results and the intrinsic free-magnon prediction arise from a combination of statistics mismatch at low $T$ (a shortcoming of LLG simulations), and genuine many-body effects at higher temperatures that our approach succeeds at capturing. 

On the technical side, which is the main focus of our work, 
our results demonstrate that semiclassical spin dynamics provide a powerful and versatile framework to study thermal Hall transport. 
In particular, they allow access to large system sizes and long simulation
times, naturally incorporate nonlinear effects beyond the harmonic
approximation, and avoid the exponential growth of the
Hilbert space inherent to full quantum approaches. 
Further developments could include a systematic study of disorder effects in thermal transport, which are known to generate important corrections through skew-scattering and side-jump mechanisms and can significantly modify the thermal Hall response~\cite{mangeolle2025extrinsic}. Another relevant extension, as discussed above, will be the implementation of colored noise to emulate Bose statistics within the LLG dynamics~\cite{barker2019semiquantum}, thereby improving the description of low-temperature transport. While quantum heat bath approaches have been proposed in the literature \cite{bergqvist2018, weber2025}, their consistent application in interacting spin systems remains technically challenging and yet to be understood in the context of transport calculations. Another interesting direction will be the study of coupled spin-phonon systems, the latter of which can also be treated semiclassically. 

The methodology we benchmarked and used here, namely extracting the thermal Hall response from dynamical correlation
functions computed from semiclassical spin dynamics, is applicable not only to transverse thermal transport, but can also be utilized to study other (especially, Hall-like) transport phenomena such as the transverse spin Seebeck effect. Because the method is unbiased and agnostic to a system's phase, it can be applied to a wide range of magnetic systems, ordered or liquid, frustrated or not, strongly fluctuating or weakly interacting. Finally, it would also be very worthwhile to go beyond the linear response framework and perform non-equilibrium simulations with an explicitly imposed temperature gradient as a direct path to extract transport coefficients~\cite{PhysRevB.105.104405}.
Such approaches would not only provide an independent
validation of linear-response calculations but also permit the study of out of equilibrium effects. 

\vspace{1cm}
\section{Acknowledgements} 
We would like to thank Joji Nasu and Shy Zhang for helpful discussions, and Ronan Sangouard for assistance with optimizing the code. A.M. acknowledges funding from the Deutsche Forschungsgemeinschaft (DFG, German Research Foundation) through TRR 173-268565370 (Project B13), and Project No.~504261060 (Emmy Noether Programme), and from the Dynamics and Topology Center (TopDyn) funded by the State of Rhineland-Palatinate. JK acknowledges support from the Deutsche Forschungsgemeinschaft (DFG, German Research Foundation) under Germany’s Excellence Strategy (EXC–2111–390814868 and ct.qmat EXC-2147-390858490),
and DFG Grants No. KN1254/1-2, KN1254/2-1 TRR 360 - 492547816 and SFB 1143 (project-id 247310070), as well as the Munich Quantum Valley, which is supported by the Bavarian state government with funds from the Hightech Agenda Bayern Plus. JK thanks the Keck foundation for support.

\bibliographystyle{apsrev4-2}
\bibliography{myrefs}

@phdthesis{carnahan2022spin,
  title={Spin Fluctuation-Induced Hall Effects in Two-Dimensional Magnets},
  author={Carnahan, Caitlin},
  year={2022},
  school={Carnegie Mellon University}
}

@book{auerbach2012interacting,
  title={Interacting electrons and quantum magnetism},
  author={Auerbach, Assa},
  year={2012},
  publisher={Springer Science \& Business Media}
}

@article{zhang2024thermal,
  title={Thermal Hall effects in quantum magnets},
  author={Zhang, Xiao-Tian and Gao, Yong Hao and Chen, Gang},
  journal={Physics Reports},
  volume={1070},
  pages={1--59},
  year={2024},
  publisher={Elsevier}
}

@article{nomura2012cross,
  title={Cross-correlated responses of topological superconductors and superfluids},
  author={Nomura, Kentaro and Ryu, Shinsei and Furusaki, Akira and Nagaosa, Naoto},
  journal={Physical review letters},
  volume={108},
  number={2},
  pages={026802},
  year={2012},
  publisher={APS}
}

@article{carnahan_thermal_2021,
  title     = {Thermal Hall effect of chiral spin fluctuations},
  author    = {Carnahan, Caitlin and Zhang, Yinhan and Xiao, Di},
  journal   = {Phys. Rev. B},
  volume    = {103},
  issue     = {22},
  pages     = {224419},
  numpages  = {6},
  year      = {2021},
  month     = {Jun},
  publisher = {American Physical Society},
  doi       = {10.1103/PhysRevB.103.224419},
  url       = {https://link.aps.org/doi/10.1103/PhysRevB.103.224419}
}

@article{yang_generalizations_1980,
  title     = {Generalizations of classical Poisson brackets to include spin},
  author    = {Yang, Kuo-Ho and Hirschfelder, Joseph O.},
  journal   = {Phys. Rev. A},
  volume    = {22},
  issue     = {5},
  pages     = {1814--1816},
  numpages  = {0},
  year      = {1980},
  month     = {Nov},
  publisher = {American Physical Society},
  doi       = {10.1103/PhysRevA.22.1814},
  url       = {https://link.aps.org/doi/10.1103/PhysRevA.22.1814}
}

@article{luttinger_theory_1964,
  title     = {Theory of Thermal Transport Coefficients},
  author    = {Luttinger, J. M.},
  journal   = {Phys. Rev.},
  volume    = {135},
  issue     = {6A},
  pages     = {A1505--A1514},
  numpages  = {0},
  year      = {1964},
  month     = {Sep},
  publisher = {American Physical Society},
  doi       = {10.1103/PhysRev.135.A1505},
  url       = {https://link.aps.org/doi/10.1103/PhysRev.135.A1505}
}

@article{qin_em,
  title     = {Berry curvature and the phonon Hall effect},
  author    = {Qin, Tao and Zhou, Jianhui and Shi, Junren},
  journal   = {Phys. Rev. B},
  volume    = {86},
  issue     = {10},
  pages     = {104305},
  numpages  = {9},
  year      = {2012},
  month     = {Sep},
  publisher = {American Physical Society},
  doi       = {10.1103/PhysRevB.86.104305},
  url       = {https://link.aps.org/doi/10.1103/PhysRevB.86.104305}
}

@article{qinniushi2011,
  title = {Energy Magnetization and the Thermal Hall Effect},
  author = {Qin, Tao and Niu, Qian and Shi, Junren},
  journal = {Phys. Rev. Lett.},
  volume = {107},
  issue = {23},
  pages = {236601},
  numpages = {5},
  year = {2011},
  month = {Nov},
  publisher = {American Physical Society},
  doi = {10.1103/PhysRevLett.107.236601},
  url = {https://link.aps.org/doi/10.1103/PhysRevLett.107.236601}
}

@article{Kim2024,
  title     = {Thermal Hall effects due to topological spin fluctuations in YMnO3},
  volume    = {15},
  issn      = {2041-1723},
  url       = {http://dx.doi.org/10.1038/s41467-023-44448-9},
  doi       = {10.1038/s41467-023-44448-9},
  number    = {1},
  journal   = {Nature Communications},
  publisher = {Springer Science and Business Media LLC},
  author    = {Kim,  Ha-Leem and Saito,  Takuma and Yang,  Heejun and Ishizuka,  Hiroaki and Coak,  Matthew John and Lee,  Jun Han and Sim,  Hasung and Oh,  Yoon Seok and Nagaosa,  Naoto and Park,  Je-Geun},
  year      = {2024},
  month     = jan
}

@article{matsumoto_rotational_2011,
  title     = {Theoretical Prediction of a Rotating Magnon Wave Packet in Ferromagnets},
  author    = {Matsumoto, Ryo and Murakami, Shuichi},
  journal   = {Phys. Rev. Lett.},
  volume    = {106},
  issue     = {19},
  pages     = {197202},
  numpages  = {4},
  year      = {2011},
  month     = {May},
  publisher = {American Physical Society},
  doi       = {10.1103/PhysRevLett.106.197202},
  url       = {https://link.aps.org/doi/10.1103/PhysRevLett.106.197202}
}

@article{mcclarty_topological_magnons_2018,
  title     = {Topological magnons in Kitaev magnets at high fields},
  author    = {McClarty, P. A. and Dong, X.-Y. and Gohlke, M. and Rau, J. G. and Pollmann, F. and Moessner, R. and Penc, K.},
  journal   = {Phys. Rev. B},
  volume    = {98},
  issue     = {6},
  pages     = {060404},
  numpages  = {6},
  year      = {2018},
  month     = {Aug},
  publisher = {American Physical Society},
  doi       = {10.1103/PhysRevB.98.060404},
  url       = {https://link.aps.org/doi/10.1103/PhysRevB.98.060404}
}

@article{Evans2014,
  title     = {Atomistic spin model simulations of magnetic nanomaterials},
  volume    = {26},
  issn      = {1361-648X},
  url       = {http://dx.doi.org/10.1088/0953-8984/26/10/103202},
  doi       = {10.1088/0953-8984/26/10/103202},
  number    = {10},
  journal   = {Journal of Physics: Condensed Matter},
  publisher = {IOP Publishing},
  author    = {Evans,  R F L and Fan,  W J and Chureemart,  P and Ostler,  T A and Ellis,  M O A and Chantrell,  R W},
  year      = {2014},
  month     = feb,
  pages     = {103202}
}

@article{ellis_midpoint_2012,
  title     = {Classical spin model of the relaxation dynamics of rare-earth doped permalloy},
  author    = {Ellis, M. O. A. and Ostler, T. A. and Chantrell, R. W.},
  journal   = {Phys. Rev. B},
  volume    = {86},
  issue     = {17},
  pages     = {174418},
  numpages  = {9},
  year      = {2012},
  month     = {Nov},
  publisher = {American Physical Society},
  doi       = {10.1103/PhysRevB.86.174418},
  url       = {https://link.aps.org/doi/10.1103/PhysRevB.86.174418}
}

@article{mook_spin_2016,
  title     = {Spin dynamics simulations of topological magnon insulators: From transverse current correlation functions to the family of magnon Hall effects},
  author    = {Mook, Alexander and Henk, J\"urgen and Mertig, Ingrid},
  journal   = {Phys. Rev. B},
  volume    = {94},
  issue     = {17},
  pages     = {174444},
  numpages  = {13},
  year      = {2016},
  month     = {Nov},
  publisher = {American Physical Society},
  doi       = {10.1103/PhysRevB.94.174444},
  url       = {https://link.aps.org/doi/10.1103/PhysRevB.94.174444}
}

@article{Smrcka_1977,
doi = {10.1088/0022-3719/10/12/021},
url = {https://dx.doi.org/10.1088/0022-3719/10/12/021},
year = {1977},
month = {jun},
publisher = {},
volume = {10},
number = {12},
pages = {2153},
author = {L Smrcka and P Streda},
title = {Transport coefficients in strong magnetic fields},
journal = {Journal of Physics C: Solid State Physics}
}

@article{cooper1997,
  title = {Thermoelectric response of an interacting two-dimensional electron gas in a quantizing magnetic field},
  author = {Cooper, N. R. and Halperin, B. I. and Ruzin, I. M.},
  journal = {Phys. Rev. B},
  volume = {55},
  issue = {4},
  pages = {2344--2359},
  numpages = {0},
  year = {1997},
  month = {Jan},
  publisher = {American Physical Society},
  doi = {10.1103/PhysRevB.55.2344},
  url = {https://link.aps.org/doi/10.1103/PhysRevB.55.2344}
}

@article{thermoEM2020,
  title = {Thermodynamics of energy magnetization},
  author = {Zhang, Yinhan and Gao, Yang and Xiao, Di},
  journal = {Phys. Rev. B},
  volume = {102},
  issue = {23},
  pages = {235161},
  numpages = {5},
  year = {2020},
  month = {Dec},
  publisher = {American Physical Society},
  doi = {10.1103/PhysRevB.102.235161},
  url = {https://link.aps.org/doi/10.1103/PhysRevB.102.235161}
}

@article{starpaper,
  title = {Quantum kinetic equation and thermal conductivity tensor for bosons},
  author = {Mangeolle, L\'eo and Savary, Lucile and Balents, Leon},
  journal = {Phys. Rev. B},
  volume = {109},
  issue = {23},
  pages = {235137},
  numpages = {18},
  year = {2024},
  month = {Jun},
  publisher = {American Physical Society},
  doi = {10.1103/PhysRevB.109.235137},
  url = {https://link.aps.org/doi/10.1103/PhysRevB.109.235137}
}

@article{sumiyoshi2013quantum,
  title={Quantum thermal Hall effect in a time-reversal-symmetry-broken topological superconductor in two dimensions: Approach from bulk calculations},
  author={Sumiyoshi, Hiroaki and Fujimoto, Satoshi},
  journal={Journal of the Physical Society of Japan},
  volume={82},
  number={2},
  pages={023602},
  year={2013},
  publisher={The Physical Society of Japan}
}

@article{kane1997quantized,
  title={Quantized thermal transport in the fractional quantum Hall effect},
  author={Kane, CL and Fisher, Matthew PA},
  journal={Physical Review B},
  volume={55},
  number={23},
  pages={15832},
  year={1997},
  publisher={APS}
}

@article{kitaev2006anyons,
  title={Anyons in an exactly solved model and beyond},
  author={Kitaev, Alexei},
  journal={Annals of Physics},
  volume={321},
  number={1},
  pages={2--111},
  year={2006},
  publisher={Elsevier}
}

@article{mangeolle2025extrinsic,
  title={Extrinsic contribution to bosonic thermal Hall transport},
  author={Mangeolle, L{\'e}o and Knolle, Johannes},
  journal={arXiv preprint arXiv:2505.09741},
  year={2025}
}

@article{shindou2014,
  title = {Thermal Hall effect of magnons in magnets with dipolar interaction},
  author = {Matsumoto, Ryo and Shindou, Ryuichi and Murakami, Shuichi},
  journal = {Phys. Rev. B},
  volume = {89},
  issue = {5},
  pages = {054420},
  numpages = {12},
  year = {2014},
  month = {Feb},
  publisher = {American Physical Society},
  doi = {10.1103/PhysRevB.89.054420},
  url = {https://link.aps.org/doi/10.1103/PhysRevB.89.054420}
}

@article{chern2020magnetic,
  title={Magnetic field induced competing phases in spin-orbital entangled Kitaev magnets},
  author={Chern, Li Ern and Kaneko, Ryui and Lee, Hyun-Yong and Kim, Yong Baek},
  journal={Physical Review Research},
  volume={2},
  number={1},
  pages={013014},
  year={2020},
  publisher={APS}
}

@article{knolle2014dynamics,
  title={Dynamics of a two-dimensional quantum spin liquid: Signatures of emergent Majorana fermions and fluxes},
  author={Knolle, Johannes and Kovrizhin, DL and Chalker, JT and Moessner, Roderich},
  journal={Physical Review Letters},
  volume={112},
  number={20},
  pages={207203},
  year={2014},
  publisher={APS}
}

@article{koyama2024thermal,
  title={Thermal Hall effect incorporating magnon damping in localized spin systems},
  author={Koyama, Shinnosuke and Nasu, Joji},
  journal={Physical Review B},
  volume={109},
  number={17},
  pages={174442},
  year={2024},
  publisher={APS}
}

@article{dimos2025thermal,
  title={Thermal Hall effect of magnons from many-body skew scattering},
  author={Chatzichrysafis, Dimos and Mook, Alexander},
  journal={Physical Review B},
  volume={111},
  number={13},
  pages={134405},
  year={2025},
  publisher={APS}
}

@article{side-jump,
  title={Extrinsic contribution to bosonic thermal Hall transport},
  author={Mangeolle, L{\'e}o and Knolle, Johannes},
  journal={arXiv preprint arXiv:2505.09741},
  year={2025}
}

@article{mcclarty2022topological,
  title={Topological magnons: A review},
  author={McClarty, Paul A},
  journal={Annual Review of Condensed Matter Physics},
  volume={13},
  number={1},
  pages={171--190},
  year={2022},
  publisher={Annual Reviews}
}

@article{daquino_symplectic,
title = {Geometrical integration of Landau–Lifshitz–Gilbert equation based on the mid-point rule},
journal = {Journal of Computational Physics},
volume = {209},
number = {2},
pages = {730-753},
year = {2005},
issn = {0021-9991},
doi = {https://doi.org/10.1016/j.jcp.2005.04.001},
url = {https://www.sciencedirect.com/science/article/pii/S002199910500197X},
author = {Massimiliano d’Aquino and Claudio Serpico and Giovanni Miano},
}

@article{Mentink_2010,
doi = {10.1088/0953-8984/22/17/176001},
url = {https://dx.doi.org/10.1088/0953-8984/22/17/176001},
year = {2010},
month = {apr},
publisher = {},
volume = {22},
number = {17},
pages = {176001},
author = {Mentink, J H and Tretyakov, M V and Fasolino, A and Katsnelson, M I and Rasing, Th},
title = {Stable and fast semi-implicit integration of the stochastic Landau–Lifshitz equation},
journal = {Journal of Physics: Condensed Matter},
}

@article{gilbert2004phenomenological,
  title={A phenomenological theory of damping in ferromagnetic materials},
  author={Gilbert, Thomas L},
  journal={IEEE transactions on magnetics},
  volume={40},
  number={6},
  pages={3443--3449},
  year={2004},
  publisher={IEEE}
}

@book{eriksson2017atomistic,
  title={Atomistic spin dynamics: foundations and applications},
  author={Eriksson, Olle and Bergman, Anders and Bergqvist, Lars and Hellsvik, Johan},
  year={2017},
  publisher={Oxford University Press}
}

@book{pottier2009nonequilibrium,
  title={Nonequilibrium statistical physics: linear irreversible processes},
  author={Pottier, No{\"e}lle},
  year={2009},
  publisher={Oxford University Press}
}

@article{barker2019semiquantum,
  title={Semiquantum thermodynamics of complex ferrimagnets},
  author={Barker, Joseph and Bauer, Gerrit EW},
  journal={Physical Review B},
  volume={100},
  number={14},
  pages={140401},
  year={2019},
  publisher={APS}
}

@article{tang2019quantized,
  title={Quantized thermal Hall conductance from edge current calculations in lattice models},
  author={Tang, Wei and Xie, XC and Wang, Lei and Tu, Hong-Hao},
  journal={arXiv preprint arXiv:1905.12475},
  year={2019}
}

@article{Baskaran2008,
  title = {Spin-$S$ Kitaev model: Classical ground states, order from disorder, and exact correlation functions},
  author = {Baskaran, G. and Sen, Diptiman and Shankar, R.},
  journal = {Phys. Rev. B},
  volume = {78},
  issue = {11},
  pages = {115116},
  numpages = {8},
  year = {2008},
  month = {Sep},
  publisher = {American Physical Society},
  doi = {10.1103/PhysRevB.78.115116},
  url = {https://link.aps.org/doi/10.1103/PhysRevB.78.115116}
}

@book{han2017skyrmions,
  title={Skyrmions in condensed matter},
  author={Han, Jung Hoon},
  volume={278},
  year={2017},
  publisher={Springer}
}

@book{frenkel2002,
  author    = {Frenkel, Daan and Smit, Berend},
  title     = {Understanding Molecular Simulation: From Algorithms to Applications},
  edition   = {2nd},
  year      = {2002},
  publisher = {Academic Press},
  address   = {San Diego},
  series    = {Computational Science Series},
  isbn      = {9780122673511}
}

@phdthesis{thesishellsvik,
  author       = {Hellsvik, Johan},
  title        = {Atomistic Spin Dynamics, Theory and Applications},
  school       = {University of Uppsala},
  year         = {2010},
  address      = {Uppsala},
}

@article{joy2022,
  title = {Dynamics of Visons and Thermal Hall Effect in Perturbed Kitaev Models},
  author = {Joy, Aprem P. and Rosch, Achim},
  journal = {Phys. Rev. X},
  volume = {12},
  issue = {4},
  pages = {041004},
  numpages = {19},
  year = {2022},
  month = {Oct},
  publisher = {American Physical Society},
  doi = {10.1103/PhysRevX.12.041004},
  url = {https://link.aps.org/doi/10.1103/PhysRevX.12.041004}
}

@article{samarkoon2018,
  title = {Classical and quantum spin dynamics of the honeycomb $\mathrm{\ensuremath{\Gamma}}$ model},
  author = {Samarakoon, Anjana M. and Wachtel, Gideon and Yamaji, Youhei and Tennant, D. A. and Batista, Cristian D. and Kim, Yong Baek},
  journal = {Phys. Rev. B},
  volume = {98},
  issue = {4},
  pages = {045121},
  numpages = {16},
  year = {2018},
  month = {Jul},
  publisher = {American Physical Society},
  doi = {10.1103/PhysRevB.98.045121},
  url = {https://link.aps.org/doi/10.1103/PhysRevB.98.045121}
}

@article{Czajka2022,
  title = {Planar thermal Hall effect of topological bosons in the Kitaev magnet α-RuCl3},
  volume = {22},
  ISSN = {1476-4660},
  url = {http://dx.doi.org/10.1038/s41563-022-01397-w},
  DOI = {10.1038/s41563-022-01397-w},
  number = {1},
  journal = {Nature Materials},
  publisher = {Springer Science and Business Media LLC},
  author = {Czajka,  Peter and Gao,  Tong and Hirschberger,  Max and Lampen-Kelley,  Paula and Banerjee,  Arnab and Quirk,  Nicholas and Mandrus,  David G. and Nagler,  Stephen E. and Ong,  N. P.},
  year = {2022},
  month = nov,
  pages = {36–41}
}

@misc{kim2025,
      title={Emulation of quantum correlations by classical dynamics in a spin-1/2 Heisenberg chain}, 
      author={Chaebin Kim and Martin Mourigal},
      year={2025},
      eprint={2503.19975},
      archivePrefix={arXiv},
      primaryClass={cond-mat.str-el},
      url={https://arxiv.org/abs/2503.19975}, 
}

@article{franke2022,
  title = {Thermal spin dynamics of Kitaev magnets: Scattering continua and magnetic field induced phases within a stochastic semiclassical approach},
  author = {Franke, Oliver and C\ifmmode \u{a}\else \u{a}\fi{}lug\ifmmode \u{a}\else \u{a}\fi{}ru, Dumitru and Nunnenkamp, Andreas and Knolle, Johannes},
  journal = {Phys. Rev. B},
  volume = {106},
  issue = {17},
  pages = {174428},
  numpages = {10},
  year = {2022},
  month = {Nov},
  publisher = {American Physical Society},
  doi = {10.1103/PhysRevB.106.174428},
  url = {https://link.aps.org/doi/10.1103/PhysRevB.106.174428}
}

@article{koyama2024,
  title = {Thermal Hall effect incorporating magnon damping in localized spin systems},
  author = {Koyama, Shinnosuke and Nasu, Joji},
  journal = {Phys. Rev. B},
  volume = {109},
  issue = {17},
  pages = {174442},
  numpages = {23},
  year = {2024},
  month = {May},
  publisher = {American Physical Society},
  doi = {10.1103/PhysRevB.109.174442},
  url = {https://link.aps.org/doi/10.1103/PhysRevB.109.174442}
}

@article{matsumoto2014,
  title = {Thermal Hall effect of magnons in magnets with dipolar interaction},
  author = {Matsumoto, Ryo and Shindou, Ryuichi and Murakami, Shuichi},
  journal = {Phys. Rev. B},
  volume = {89},
  issue = {5},
  pages = {054420},
  numpages = {12},
  year = {2014},
  month = {Feb},
  publisher = {American Physical Society},
  doi = {10.1103/PhysRevB.89.054420},
  url = {https://link.aps.org/doi/10.1103/PhysRevB.89.054420}
}

@article{vinkler2018approximately,
  title={Approximately quantized thermal Hall effect of chiral liquids coupled to phonons},
  author={Vinkler-Aviv, Yuval and Rosch, Achim},
  journal={Physical Review X},
  volume={8},
  number={3},
  pages={031032},
  year={2018},
  publisher={APS}
}

@article{chernyshev2009spin,
  title={Spin waves in a triangular lattice antiferromagnet: Decays, spectrum renormalization, and singularities},
  author={Chernyshev, AL and Zhitomirsky, ME},
  journal={Physical Review B—Condensed Matter and Materials Physics},
  volume={79},
  number={14},
  pages={144416},
  year={2009},
  publisher={APS}
}

@article{RevModPhys.85.219,
  title = {Colloquium: Spontaneous magnon decays},
  author = {Zhitomirsky, M. E. and Chernyshev, A. L.},
  journal = {Rev. Mod. Phys.},
  volume = {85},
  issue = {1},
  pages = {219--242},
  numpages = {0},
  year = {2013},
  month = {Jan},
  publisher = {American Physical Society},
  doi = {10.1103/RevModPhys.85.219},
  url = {https://link.aps.org/doi/10.1103/RevModPhys.85.219}
}

@article{PhysRevB.91.125413,
  title = {Thermal Hall effect of spins in a paramagnet},
  author = {Lee, Hyunyong and Han, Jung Hoon and Lee, Patrick A.},
  journal = {Phys. Rev. B},
  volume = {91},
  issue = {12},
  pages = {125413},
  numpages = {8},
  year = {2015},
  month = {Mar},
  publisher = {American Physical Society},
  doi = {10.1103/PhysRevB.91.125413},
  url = {https://link.aps.org/doi/10.1103/PhysRevB.91.125413}
}

@article{mangeolle2022thermal,
  title={Thermal conductivity and theory of inelastic scattering of phonons by collective fluctuations},
  author={Mangeolle, L{\'e}o and Balents, Leon and Savary, Lucile},
  journal={Physical Review B},
  volume={106},
  number={24},
  pages={245139},
  year={2022},
  publisher={APS}
}

@article{mangeolle2022phonon,
  title={Phonon thermal Hall conductivity from scattering with collective fluctuations},
  author={Mangeolle, L{\'e}o and Balents, Leon and Savary, Lucile},
  journal={Physical Review X},
  volume={12},
  number={4},
  pages={041031},
  year={2022},
  publisher={APS}
}

@article{okubo2025thermal,
  title={Thermal Hall transport in Kitaev spin liquids},
  author={Okubo, Tsuyoshi and Nasu, Joji and Misawa, Takahiro and Motome, Yukitoshi},
  journal={arXiv preprint arXiv:2507.16558},
  year={2025}
}

@article{ye2018quantization,
  title={Quantization of the thermal Hall conductivity at small Hall angles},
  author={Ye, Mengxing and Hal{\'a}sz, G{\'a}bor B and Savary, Lucile and Balents, Leon},
  journal={Physical review letters},
  volume={121},
  number={14},
  pages={147201},
  year={2018},
  publisher={APS}
}

@book{ziman2001electrons,
  title={Electrons and phonons: the theory of transport phenomena in solids},
  author={Ziman, John M},
  year={2001},
  publisher={Oxford university press}
}

@article{PhysRevB.107.L220406,
  title = {Thermal Hall conductivity near field-suppressed magnetic order in a Kitaev-Heisenberg model},
  author = {Kumar, Aman and Tripathi, Vikram},
  journal = {Phys. Rev. B},
  volume = {107},
  issue = {22},
  pages = {L220406},
  numpages = {5},
  year = {2023},
  month = {Jun},
  publisher = {American Physical Society},
  doi = {10.1103/PhysRevB.107.L220406},
  url = {https://link.aps.org/doi/10.1103/PhysRevB.107.L220406}
}

@article{PhysRevB.75.214305,
  title = {Thermal conductivity of a classical one-dimensional spin-phonon system},
  author = {Savin, A. V. and Tsironis, G. P. and Zotos, X.},
  journal = {Phys. Rev. B},
  volume = {75},
  issue = {21},
  pages = {214305},
  numpages = {16},
  year = {2007},
  month = {Jun},
  publisher = {American Physical Society},
  doi = {10.1103/PhysRevB.75.214305},
  url = {https://link.aps.org/doi/10.1103/PhysRevB.75.214305}
}

@article{PhysRevB.92.134305,
  title = {Spin and thermal conductivity in a classical disordered spin chain},
  author = {Jencic, B. and Prelovsek, P.},
  journal = {Phys. Rev. B},
  volume = {92},
  issue = {13},
  pages = {134305},
  numpages = {5},
  year = {2015},
  month = {Oct},
  publisher = {American Physical Society},
  doi = {10.1103/PhysRevB.92.134305},
  url = {https://link.aps.org/doi/10.1103/PhysRevB.92.134305}
}

@article{PhysRevB.100.144416,
  title = {Effects of magnetic anisotropy on spin and thermal transport in classical antiferromagnets on the square lattice},
  author = {Aoyama, Kazushi and Kawamura, Hikaru},
  journal = {Phys. Rev. B},
  volume = {100},
  issue = {14},
  pages = {144416},
  numpages = {19},
  year = {2019},
  month = {Oct},
  publisher = {American Physical Society},
  doi = {10.1103/PhysRevB.100.144416},
  url = {https://link.aps.org/doi/10.1103/PhysRevB.100.144416}
}

@article{PhysRevB.106.224407,
  title = {Spin and thermal transport and critical phenomena in three-dimensional antiferromagnets},
  author = {Aoyama, Kazushi},
  journal = {Phys. Rev. B},
  volume = {106},
  issue = {22},
  pages = {224407},
  numpages = {15},
  year = {2022},
  month = {Dec},
  publisher = {American Physical Society},
  doi = {10.1103/PhysRevB.106.224407},
  url = {https://link.aps.org/doi/10.1103/PhysRevB.106.224407}
}

@article{PhysRevB.105.104405,
  title = {Thermal conductivity of square ice},
  author = {Sutcliffe, Ruairidh and Rau, Jeffrey G.},
  journal = {Phys. Rev. B},
  volume = {105},
  issue = {10},
  pages = {104405},
  numpages = {19},
  year = {2022},
  month = {Mar},
  publisher = {American Physical Society},
  doi = {10.1103/PhysRevB.105.104405},
  url = {https://link.aps.org/doi/10.1103/PhysRevB.105.104405}
}

@article{PhysRevB.95.020401,
  title = {Magnon transport in noncollinear spin textures: Anisotropies and topological magnon Hall effects},
  author = {Mook, Alexander and G\"obel, B\"orge and Henk, J\"urgen and Mertig, Ingrid},
  journal = {Phys. Rev. B},
  volume = {95},
  issue = {2},
  pages = {020401},
  numpages = {5},
  year = {2017},
  month = {Jan},
  publisher = {American Physical Society},
  doi = {10.1103/PhysRevB.95.020401},
  url = {https://link.aps.org/doi/10.1103/PhysRevB.95.020401}
}

@article{PhysRevB.98.060405,
  title = {Topological excitations in the ferromagnetic Kitaev-Heisenberg model},
  author = {Joshi, Darshan G.},
  journal = {Phys. Rev. B},
  volume = {98},
  issue = {6},
  pages = {060405},
  numpages = {5},
  year = {2018},
  month = {Aug},
  publisher = {American Physical Society},
  doi = {10.1103/PhysRevB.98.060405},
  url = {https://link.aps.org/doi/10.1103/PhysRevB.98.060405}
}

@book{tritt2005thermal,
  title={Thermal conductivity: theory, properties, and applications},
  author={Tritt, Terry M},
  year={2005},
  publisher={Springer}
}

@article{cookmeyer2018,
  title = {Spin-wave analysis of the low-temperature thermal Hall effect in the candidate Kitaev spin liquid $\ensuremath{\alpha}\ensuremath{-}{RuCl}_{3}$},
  author = {Cookmeyer, Tessa and Moore, Joel E.},
  journal = {Phys. Rev. B},
  volume = {98},
  issue = {6},
  pages = {060412},
  numpages = {5},
  year = {2018},
  month = {Aug},
  publisher = {American Physical Society},
  doi = {10.1103/PhysRevB.98.060412},
  url = {https://link.aps.org/doi/10.1103/PhysRevB.98.060412}
}

@article{chern2021,
  title = {Sign Structure of Thermal Hall Conductivity and Topological Magnons for In-Plane Field Polarized Kitaev Magnets},
  author = {Chern, Li Ern and Zhang, Emily Z. and Kim, Yong Baek},
  journal = {Phys. Rev. Lett.},
  volume = {126},
  issue = {14},
  pages = {147201},
  numpages = {6},
  year = {2021},
  month = {Apr},
  publisher = {American Physical Society},
  doi = {10.1103/PhysRevLett.126.147201},
  url = {https://link.aps.org/doi/10.1103/PhysRevLett.126.147201}
}

@article{bergqvist2018,
  title = {Realistic finite temperature simulations of magnetic systems using quantum statistics},
  author = {Bergqvist, Lars and Bergman, Anders},
  journal = {Phys. Rev. Mater.},
  volume = {2},
  issue = {1},
  pages = {013802},
  numpages = {9},
  year = {2018},
  month = {Jan},
  publisher = {American Physical Society},
  doi = {10.1103/PhysRevMaterials.2.013802},
  url = {https://link.aps.org/doi/10.1103/PhysRevMaterials.2.013802}
}

@misc{weber2025,
      title={Atomistic spin dynamics with quantum colored noise}, 
      author={Fried-Conrad Weber and Felix Hartmann and Matias Bargheer and Janet Anders and Richard F. L. Evans},
      year={2025},
      eprint={2508.11315},
      archivePrefix={arXiv},
      primaryClass={cond-mat.mtrl-sci},
      url={https://arxiv.org/abs/2508.11315}, 
}

@article{hou2017,
  title = {Thermally driven topology in chiral magnets},
  author = {Hou, Wen-Tao and Yu, Jie-Xiang and Daly, Morgan and Zang, Jiadong},
  journal = {Phys. Rev. B},
  volume = {96},
  issue = {14},
  pages = {140403},
  numpages = {5},
  year = {2017},
  month = {Oct},
  publisher = {American Physical Society},
  doi = {10.1103/PhysRevB.96.140403},
  url = {https://link.aps.org/doi/10.1103/PhysRevB.96.140403}
}

@article{ezawa2011,
  title = {Compact merons and skyrmions in thin chiral magnetic films},
  author = {Ezawa, Motohiko},
  journal = {Phys. Rev. B},
  volume = {83},
  issue = {10},
  pages = {100408},
  numpages = {4},
  year = {2011},
  month = {Mar},
  publisher = {American Physical Society},
  doi = {10.1103/PhysRevB.83.100408},
  url = {https://link.aps.org/doi/10.1103/PhysRevB.83.100408}
}

@article{banerjee2014,
  title = {Enhanced Stability of Skyrmions in Two-Dimensional Chiral Magnets with Rashba Spin-Orbit Coupling},
  author = {Banerjee, Sumilan and Rowland, James and Erten, Onur and Randeria, Mohit},
  journal = {Phys. Rev. X},
  volume = {4},
  issue = {3},
  pages = {031045},
  numpages = {10},
  year = {2014},
  month = {Sep},
  publisher = {American Physical Society},
  doi = {10.1103/PhysRevX.4.031045},
  url = {https://link.aps.org/doi/10.1103/PhysRevX.4.031045}
}
\end{document}